\documentclass[journal,letterpaper,12pt,onecolumn,draftclsnofoot]{IEEEtran}
\pdfoutput=1

\usepackage[utf8]{inputenc}
\usepackage[T1]{fontenc}
\usepackage{amsmath,amssymb,amsfonts,latexsym,bm,bbm}
\usepackage{cite}
\usepackage[pdftex]{graphicx}
\usepackage[caption=false,font=footnotesize]{subfig}
\usepackage{epstopdf}
\usepackage{tabularx}
\usepackage{color}

\DeclareMathOperator*{\argmin}{arg\,min}
\DeclareMathOperator*{\argmax}{arg\,max}

\begin{document}
	\title{EP-based Joint Active User Detection and Channel Estimation for Massive Machine-Type Communications}
	\author{Jinyoup~Ahn,~\IEEEmembership{Student Member,~IEEE},~Byonghyo~Shim,~\IEEEmembership{Senior Member,~IEEE},~and~Kwang~Bok~Lee,~\IEEEmembership{Fellow,~IEEE}
	\thanks{J. Ahn, B. Shim, and K. B. Lee are with the Department of Electrical and Computer Engineering and the Institute of New Media and Communications (INMC), Seoul National University, Seoul 08826, Korea (e-mail: \{ahnjymcl,bshim,klee\}@snu.ac.kr).}
	\thanks{This work was supported by the Institute for Information and Communications Technology Promotion through Korea Government under grant 2016-0-00209.}
	\thanks{This paper was presented in part at the IEEE International Conference on Communications Workshops (ICC Workshops), 2018\cite{Ahn18ICC}.}}
		
	\maketitle
	\begin{abstract}
		Massive machine-type communication (mMTC) is a newly introduced service category in 5G wireless communication systems to support a variety of Internet-of-Things (IoT) applications. In recovering sparsely represented multi-user vectors, compressed sensing based multi-user detection (CS-MUD) can be used. CS-MUD is a feasible solution to the grant-free uplink non-orthogonal multiple access (NOMA) environments. In CS-MUD, active user detection (AUD) and channel estimation (CE) should be performed before data detection. In this paper, we propose the expectation propagation based joint AUD and CE (EP-AUD/CE) technique for mMTC networks. The expectation propagation (EP) algorithm is a Bayesian framework that approximates a computationally intractable probability distribution to an easily tractable distribution. The proposed technique finds the best approximation of the posterior distribution of the sparse channel vector. Using the approximate distribution, AUD and CE are jointly performed. We show by numerical simulations that the proposed technique substantially enhances AUD and CE performances over competing algorithms.
	\end{abstract}
	
	
	\section{Introduction}

\begin{figure}[t]
		\centering
		\includegraphics[draft=false,width=0.8\linewidth]{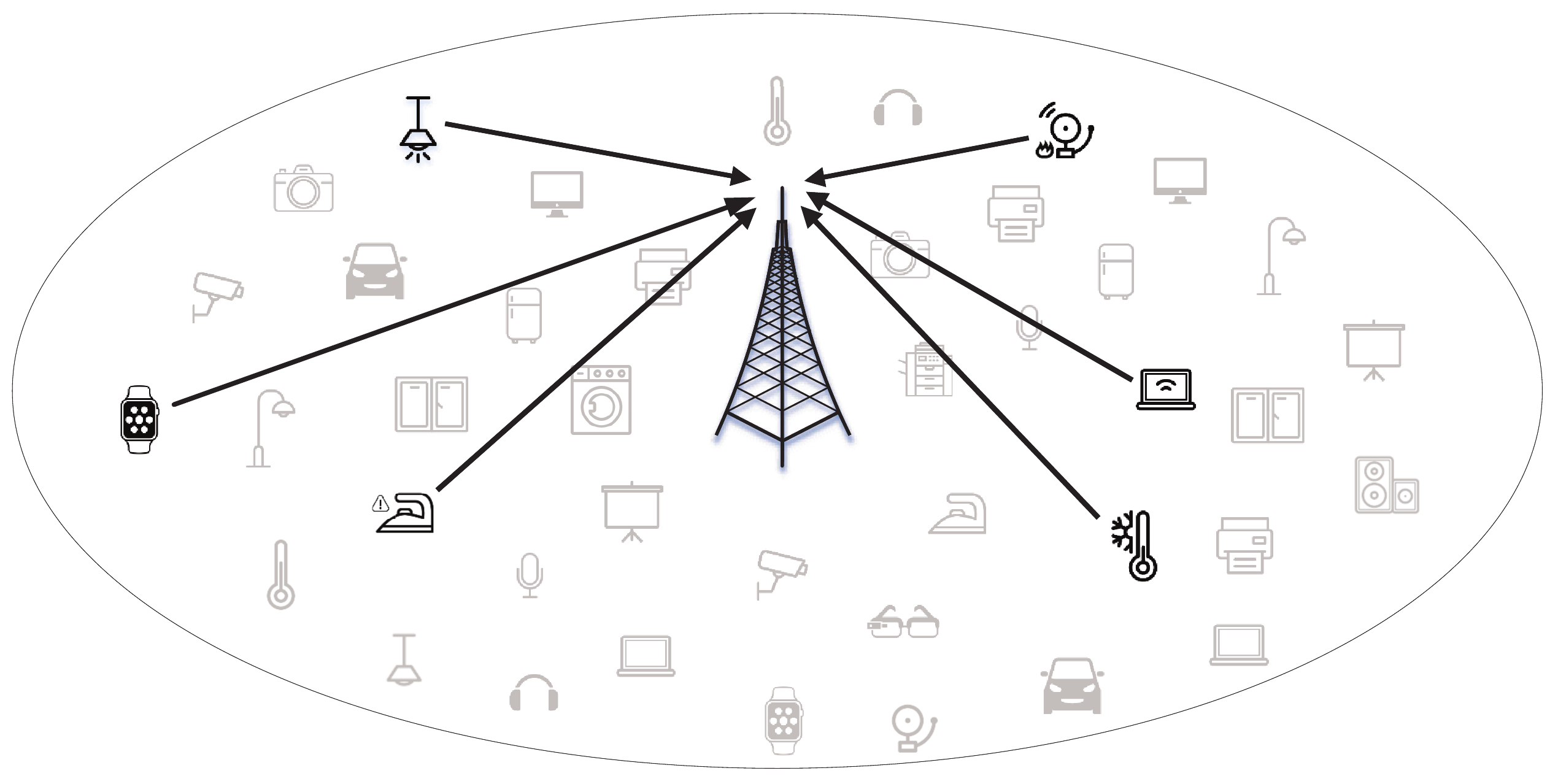}
		\caption{The sporadic uplink multiple access in a mMTC network.}
		\label{fig:mMTC}
	\end{figure}

	Recently, massive machine-type communication (mMTC) has received much attention due to its wide variety of Internet-of-Things (IoT) applications such as smart metering, factory automation, autonomous driving, surveillance, and health monitoring, to name just a few\cite{Bockelmann18}. In accordance with this trend, the International Telecommunication Union (ITU) defined mMTC as one of the key service categories for 5G wireless communications \cite{ITU15}. As illustrated in Fig.~\ref{fig:mMTC}, mMTC concerns the \textit{massive connectivity} of a large number of machine-type devices (MTDs) to the base station (BS). mMTC is distinctive from human-centric communications in the sense that the data traffic is \textit{uplink-dominated}, and devices are \textit{sporadically active} only for a short period of time to transmit \textit{short packets} with \textit{low data rates} \cite{Dawy17,Shariatmadari15}. 
	
	In the mMTC network, conventional scheduling-based multiple access schemes in which the BS allocates orthogonal time/frequency resources to each device is not relevant due to the significant signaling overhead and excessive latency caused by complicated scheduling procedure. To overcome these drawbacks, grant-free \textit{non-orthogonal multiple access} (NOMA) schemes have been proposed in recent years \cite{Shirvanimoghaddam17,Ding17}. In grant-free NOMA schemes, devices transmit data symbols in a non-orthogonal manner without relying on the granting procedures.
	
	In the mMTC network, a large portion of devices is inactive and hence does not transmit data. Thus, the transmit vector consisting of data symbols of both active and inactive devices can be readily modeled as a sparse vector. By capitalizing on the sparsity of this multi-user vector, the multi-user detection (MUD) problem can be formulated as a sparse signal recovery problem \cite{Jeong18TVT, Ahn18WCL, Shim17, Shim14, Shim12}. This type of detection scheme, called compressed sensing based multi-user detection (CS-MUD), has been a key ingredient in the grant-free uplink NOMA schemes.
	
	In recent years, several approaches have been proposed to cast the CS-MUD problem into a sparsity-aware maximum \textit{a posteriori} probability (S-MAP) detection problem. In \cite{Zhu11}, linear relaxed S-MAP detectors have been suggested. In \cite{Knoop13}, sparsity-aware successive interference cancellation (SA-SIC) has been suggested. In \cite{Barik14}, sparsity-aware sphere decoding (SA-SD) has been proposed. In \cite{Knoop14}, K-Best detection has been proposed as a variant of SA-SD. In these studies, the uplink channel state information (CSI) from the MTD to the BS is assumed to be perfectly known to the BS. In practice, however, the uplink CSI from the devices to the BS should be estimated before data detection. To address this issue, various joint active user detection (AUD) and channel estimation (CE) schemes have been proposed. Since only a few devices are active at one time, an element-wise (i.e., Hadamard) product of the binary activity pattern and the channel vector is also a sparse vector and thus compressed sensing (CS)-based technique is a good fit for the problem at hand \cite{Shim172,Chen17,Knoop16,Hannak15,Xu15}.

	One potential shortcoming in these studies is that a prior distribution of the sparse vector is not exploited. In fact, these studies are based on the non-Bayesian greedy algorithms such as the orthogonal matching pursuit (OMP) and approximate message passing (AMP) algorithms, which do not require a prior distribution of the sparse vector. In essence, these algorithms find out non-zero values based on the instantaneous correlation between the sensing matrix and the observation vector so that they might not be effective in the situation where the prior distribution is available. In this case, clearly, by exploiting the statistical distribution of the sparse vector, the performance of AUD and CE can be improved substantially.
	
	An aim of this paper is to propose a novel Bayesian joint AUD and CE technique based on the expectation propagation (EP) algorithm \cite{Opper00,Minka01,Andersen14,Hernandez15}. The EP algorithm is a Bayesian technique to approximate a computationally intractable target probability distribution to the distribution from a tractable family. By iteratively minimizing the Kullback-Leibler divergence between the target distribution and the approximate distribution via moment matching, the EP algorithm can efficiently find  a tractable approximation of the target distribution. In describing the prior distribution of activity and channel of each device, we employ the Bernoulli-Gaussian probabilistic model. The posterior distribution of user activities and channels is computationally intractable due to the discrete nature of the binary activity variables. In this work, we iteratively find the best approximation of the posterior distribution of the composite vector of user activities and channels using the EP algorithm. Using the obtained approximation, activity identification and CSI estimation of active devices are performed jointly. The data detection of active devices is then performed based on the obtained knowledge of user activities and channels. We show from numerical evaluations in realistic mMTC scenarios that the proposed technique outperforms conventional non-Bayesian greedy algorithms and other Bayesian techniques. In particular, the proposed technique performs close to the Oracle detector, an ideal detector having perfect knowledge on the user activities in the high signal-to-noise-ratio (SNR) regime.
	
	The rest of the paper is organized as follows. In Section II, we describe the system model. In Section III, we describe the joint AUD and CE problem. In Section IV, we provide the brief introduction of the EP algorithm. In Section V, we discuss the proposed EP-based joint AUD and CE method. In Section VI, we provide the simulation results and the conclusion is given in Section VII.
	
	\textit{Notation}: Scalars are denoted as lower-case letters, vectors as bold-face lower-case letters, and matrices as bold-face upper-case letters, respectively. The identity matrix and the all-zero vector of size $M$ are denoted as $\mathbf{I}_M$ and $\mathbf{0}_M$, respectively. For a matrix $\mathbf{A}$ of arbitrary size, $\mathbf{A}^{-1}$ and $\mathbf{A}^H$ denote its inverse and conjugate transpose, respectively. The distribution of a circularly symmetric complex Gaussian (CSCG) random vector $\mathbf{x}$ with mean $\mathbf{m}$ and covariance matrix $\mathbf{V}$ is denoted as $\mathcal{CN}(\mathbf{x}|\mathbf{m},\mathbf{V})$. Also, the space of complex vectors of size $M$ is denoted as $\mathbb{C}^{M}$ and the space of complex matrices of size $M \times N$ as $\mathbb{C}^{M \times N}$. The expectation operator is denoted as $E[\cdot]$ and the variance operator as $\mathrm{Var}[\cdot]$.
	
	\section{System model}
	
	\begin{figure}[t]
		\centering
		\includegraphics[draft=false,width=0.8\linewidth]{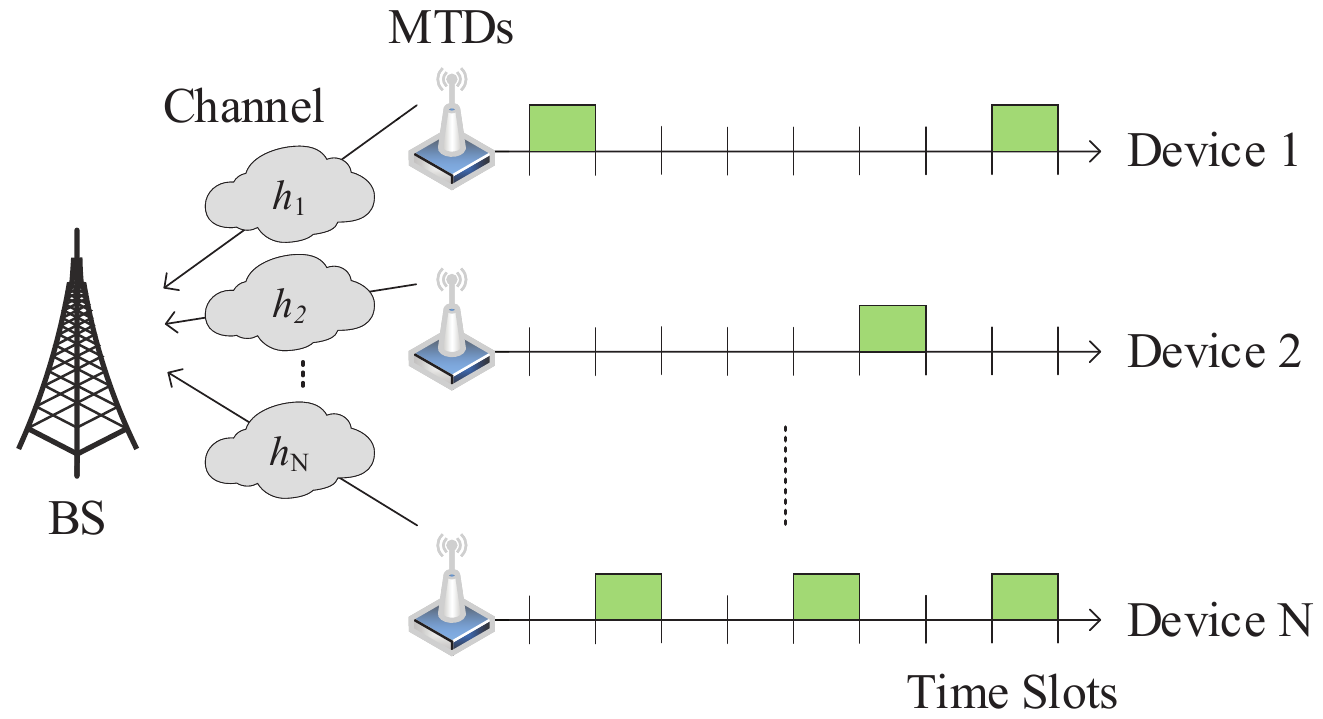}
		\caption{The mMTC uplink multiple access scenario with $N$ devices sporadically becoming active.}
		\label{fig:CS-MUD}
	\end{figure}
	
	We consider the uplink of a mMTC network where $N$ MTDs access a single BS, as shown in Fig.~\ref{fig:CS-MUD}. We assume that the BS and each device are equipped with one antenna. Each active device transmits a pilot symbol followed by $J$ data symbols (which we call a frame in the sequel). In this work, we assume that devices are synchronized in time, meaning that all devices switch their activity in the same time slot.\footnote{Since the packet size of MTD is typically very small ($10\sim100$ bytes) in mMTC environment, the impact of this assumption on the proposed scheme would be marginal.} Each device is either active or inactive in the whole interval of the frame. Also, we assume a flat fading channel model where channel remains unchanged in the entire frame. We denote the complex uplink channel coefficient from the $n$-th device to the BS by $h_n$, where $h_n$ follows the zero-mean complex-Gaussian non-dispersive independent Rayleigh fading with the variance $\alpha_n$, i.e., $h_n \sim \mathcal{CN}(h_n|0,\alpha_n)$. The variance $\alpha_n$ captures the pathloss component characterized by the each device's location. We assume that $\alpha_n$ is known at the BS. In the scenarios where devices are stationary, the pathloss can be estimated and stored at the BS as a prior information. 
	
	In order to model the sporadic traffic pattern of the mMTC network, we define the binary user activity indicator $a_n$ for the $n$-th device as
	\begin{align}\label{eqn:activity-model}
		a_n &= \left\{ \begin{array}{ll}
		1, & \textrm{if the $n$-th device is active},\\
		0, & \textrm{otherwise}.
	\end{array} \right.
	\end{align}
	The $n$-th device is active with the activity probability $p_{n}$ and the activity of each device is independent of each other. 
	
	In this setup, the input-output relationship can be described as 
	\begin{equation}\label{eqn:sys-model-basic}
		\mathbf{y}=\sum_{n=1}^{N} \mathbf{s}_n h_n a_n x_n +\mathbf{w},
	\end{equation}
	where $x_n$ is the transmit symbol of the $n$-th device, $\mathbf{s}_n \in \mathbb{C}^M$ is the spreading sequence for the $n$-th device, $\mathbf{y} \in \mathbb{C}^M$ is the measurement vector at the BS, and $\mathbf{w}$ is the independent zero-mean complex-Gaussian noise vector with the variance $\sigma_w^2$. We assume that the transmit powers of all devices are the same, i.e., $E[|x_n|^2]=\rho$. 
	
	In the general mMTC scenarios, the number of devices $N$ is larger than the number of resources used for the transmission $M$ (i.e, $M < N$). While it is in general not possible to recover the target vector in this underdetermined scenario, the theory of compressed sensing (CS) guarantees that the target vector can be recovered accurately as long as the vector is sparse and the measurement process preserves the energy of an input vector \cite{Shim17}.
	
	\section{Joint Active User Detection and Channel Estimation}
	
	\begin{figure}
		\centering
		\includegraphics[width=0.8\linewidth]{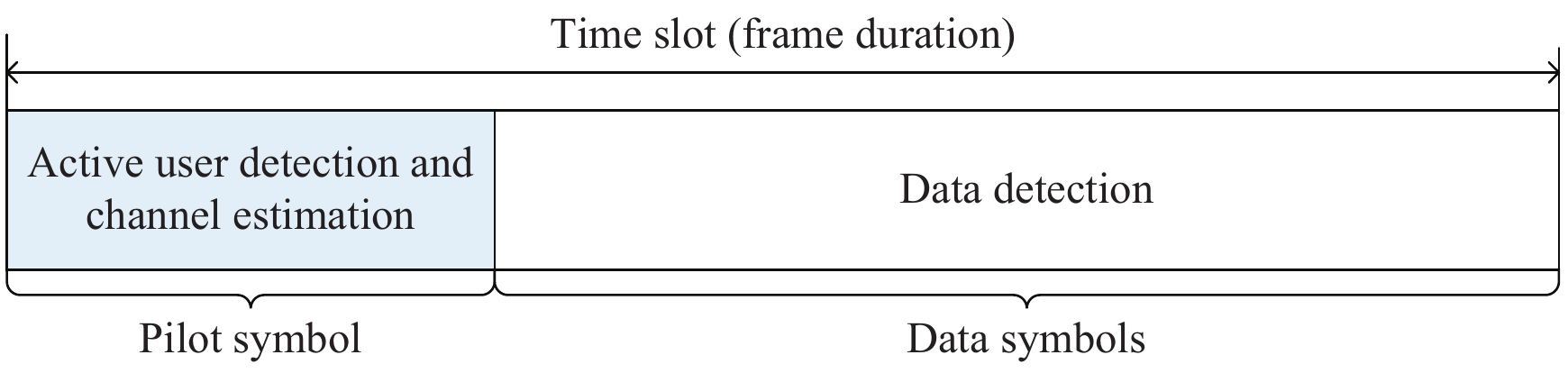}
		\caption{Two-phase grant-free multiple access protocol.}
		\label{fig:Two_Phases}
	\end{figure}

	In this work, we adopt a grant-free multiple access protocol consisting of two operational phases in each time slot (see Fig.~\ref{fig:Two_Phases}). In the first phase, each active device transmits the pilot symbol to the BS and the BS jointly detects the user activities and then estimate the channels of active devices. In the second phase, the active devices transmit $J$ data symbols to the BS and the BS decodes the data symbols using the obtained user activities and channels.
	
	In this section, we describe the joint active user detection (AUD) and channel estimation (CE). Each active device transmits the pilot symbol $x_{p,n}$ to the BS before the data transmission. The pilot measurement vector $\mathbf{y}_p$ is given by
	\begin{equation}
		\mathbf{y}_p=\sum_{n=1}^{N} \mathbf{s}_n h_n a_n x_{p,n} +\mathbf{w}_p. 
		\label{eqn:sys-model-pilot}
	\end{equation}
	Let $\bm{\phi}_n = \mathbf{s}_n x_{p,n}$ and $\mathbf{\Phi} = [\bm{\phi}_1,\dots,\bm{\phi}_N]$, then we have
	\begin{align}
		\mathbf{y}_p &=\sum_{n=1}^{N} \bm{\phi}_n a_n h_n +\mathbf{w}_p, 	\nonumber\\
		&=\mathbf{\Phi}(\mathbf{a}\circ\mathbf{h})+\mathbf{w}_p, 
		\label{eqn:sys-model-pilot2}
	\end{align}
	where $\mathbf{a}=[a_1,\dots,a_N]^T$ is the activity vector, $\mathbf{h}=[h_1,\dots,h_N]^T$ is the channel vector, and $\circ$ is the Hadamard (element-wise) product operator.
	
	Further by denoting the composite of activity vector $\mathbf{a}$ and channel vector $\mathbf{h}$ as $\mathbf{g}=\mathbf{a}\circ\mathbf{h}=\left[a_1h_1,\cdots,a_Nh_N\right]^T$, we have
	\begin{align}
		\mathbf{y}_p &=\mathbf{\Phi}\mathbf{g}+\mathbf{w}_p.
		\label{eqn:sys-model-pilot3}
	\end{align}
	
	The output of the MAP estimator maximizing the \textit{a posteriori} probability $p({\mathbf{g}}|{\mathbf{y}_p})$ is given by
	\begin{align}
		\hat{\mathbf{g}} &= \argmax\limits_{\mathbf{g} \in \mathbb{C}^N} p(\mathbf{g}|\mathbf{y}_p) \nonumber\\
		&= \argmax\limits_{\mathbf{g} \in \mathbb{C}^N} p(\mathbf{y}_p|\mathbf{g})p(\mathbf{g}),
		\label{eqn:S-MAP estimator-pre}
	\end{align}
	where $p(\mathbf{y}_p|\mathbf{g}) = \mathcal{CN}(\mathbf{y}_p|\mathbf{\Phi g},\sigma_w^2\mathbf{I})$ is the likelihood function of $\mathbf{y}_p$ given $\mathbf{g}$.
	
	When the $n$-th device is active ($a_n=1$) with an activity probability $p_{n}$ and further $p_{n}$ has a very small value for all $n$, $\mathbf{g}$ can be readily modeled as a sparse vector. Considering that the activities of devices are independent of each other, the prior distribution of $\mathbf{g}$ can be expressed as
	\begin{align}
		p(\mathbf{g})&=\prod_{n=1}^{N}\left[(1-p_n){\delta(g_n)} + p_n{\mathcal{CN}(g_n|0,\alpha_n)}\right], \label{eqn:spike-and-slab prior}
	\end{align}
	where $\delta(\cdot)$ is the Dirac delta function.
	
	From (\ref{eqn:S-MAP estimator-pre}) and (\ref{eqn:spike-and-slab prior}), we have
	\begin{align}
		\hat{\mathbf{g}} &= \argmax\limits_{\mathbf{g} \in \mathbb{C}^N} p(\mathbf{y}_p|\mathbf{g})p(\mathbf{g})  \nonumber\\
		&=\argmax\limits_{\mathbf{g} \in \mathbb{C}^N}\mathcal{CN}(\mathbf{y}_p|\mathbf{\Phi g},\sigma_w^2\mathbf{I})\prod_{n=1}^{N}\left[(1-p_n)\delta(g_n) + p_n\mathcal{CN}(g_n|0,\alpha_n)\right]. 
		\label{eqn:S-MAP estimator}
	\end{align}
	
	The goal of the MAP estimator is to find out a vector maximizing the cost function in (\ref{eqn:S-MAP estimator}). Unfortunately, when the number of devices $N$ is very large, the optimization problem in (\ref{eqn:S-MAP estimator}) is computationally intractable due to the discrete nature of the binary activity vector $\mathbf{a}$. In this situation, it is desirable to construct a tractable approximation of the target posterior distribution to solve the MAP problem, i.e., $q(\mathbf{g}) \approx p(\mathbf{y}_p|\mathbf{g})p(\mathbf{g})$. In fact, the main idea of the proposed technique is to construct a multivariate Gaussian approximation of $f(\mathbf{g}) = p(\mathbf{y}_p|\mathbf{g})p(\mathbf{g})$ and then find out the mean and variance matching to $f(\mathbf{g})$ using the expectation propagation (EP) algorithm \cite{Opper00,Minka01,Andersen14,Hernandez15}. It has been shown that with only a few numbers of iterations, the EP algorithm can achieve an accurate approximate distribution with high probability \cite{Opper00,Minka01}.
	
	\section{EP-Based Active User Detection and Channel Estimation}
	In this section, we describe the proposed EP-based joint AUD and CE method. First, we approximate the target distribution $f(\mathbf{g}) = p(\mathbf{y}_p|\mathbf{g})p(\mathbf{g})$ in \eqref{eqn:S-MAP estimator} to the Gaussian distribution $q(\mathbf{g}) = \mathcal{CN}(\mathbf{g}|\tilde{\mathbf{m}},\tilde{\mathbf{V}})$. Then, we match the mean vector $\tilde{\mathbf{m}}$ and covariance matrix $\tilde{\mathbf{V}}$ to those of the true target distribution $f(\mathbf{g})$ based on the iterative EP algorithm. 
	
	\begin{figure}
		\centering
		\includegraphics[width=\linewidth]{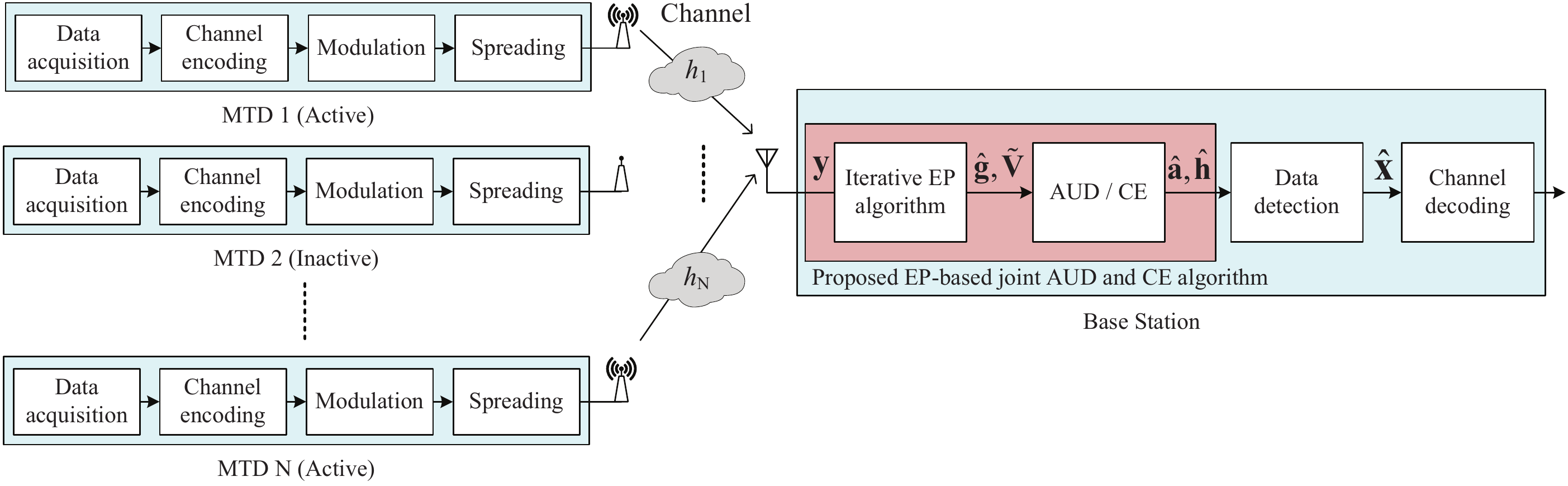}
		\caption{Block diagram of the proposed EP-based joint AUD and CE algorithm.}
		\label{fig:Block_Diagram}
	\end{figure}
	
	After the convergence, the approximate mean vector $\tilde{\mathbf{m}}$ becomes a reliable solution $\hat{\mathbf{g}}$ to the MAP estimation problem in \eqref{eqn:S-MAP estimator} and its covariance matrix is $\tilde{\mathbf{V}}$. Then, by performing the log-likelihood test on $\hat{\mathbf{g}}$, we detect the active devices and then estimate the CSI of the active devices. Finally, using the obtained knowledge of user activities and channels, data symbols of active devices are detected. Fig.~\ref{fig:Block_Diagram} depicts the block diagram of the proposed technique.
	
	\subsection{A Brief Review of Expectation Propagation}
	In this subsection, we briefly review the EP algorithm. The EP algorithm is an iterative algorithm that approximates the target probability distribution with a distribution from the exponential family $\mathcal{F}$. Suppose we have a target distribution $f(\mathbf{x})$ that can be factorized as
	\begin{equation}
		f(\mathbf{x}) = \prod_{i=1}^{I}f_i(\mathbf{x}).
	\end{equation}
	
	Using the EP algorithm, we can construct a tractable approximation of $f(\mathbf{x})$ with a distribution $q(\mathbf{x}) = \prod_{i=1}^{I}q_i(\mathbf{x})$, where $q_i(\mathbf{x}) \in \mathcal{F}$.
	
	To find the distribution $q(\mathbf{x})$ close to $f(\mathbf{x})$ from the exponential family $\mathcal{F}$, we use the Kullback-Leibler (KL) divergence criterion given by \cite{Kullback97}
	\begin{align}
		q(\mathbf{x}) = \argmin_{q'(\mathbf{x}) \in \mathcal{F}}\textrm{D}_{\textrm{KL}} \left[f(\mathbf{x}) \| q'(\mathbf{x})\right]. 
		\label{eqn:EP_KL divergence_1}
	\end{align}
	
	Because $q(\mathbf{x})$ belongs to the exponential family, the unique solution of the problem \eqref{eqn:EP_KL divergence_1} is obtained by matching the expected sufficient statistics (moments) of $f(\mathbf{x})$ and $q(\mathbf{x})$ \cite{Bishop06}. That is, parameters of $q(\mathbf{x})$ are chosen such that 
	\begin{align}
		E_{f(\mathbf{x})}[\mathbf{x}] &= E_{q(\mathbf{x})}[\mathbf{x}], \label{eqn:tilda_m}\\
		\mathrm{Var}_{f(\mathbf{x})}[\mathbf{x}] &= \mathrm{Var}_{q(\mathbf{x})}[\mathbf{x}].
		\label{eqn:tilda_V}
	\end{align}
	
	In the EP algorithm, the parameters of $q(\mathbf{x})$ satisfying \eqref{eqn:tilda_m} and \eqref{eqn:tilda_V} are obtained iteratively. In each iteration of the EP algorithm, $q_i(\mathbf{x})$ in $q(\mathbf{x})$ is replaced by $f_i(\mathbf{x})$. In other words, we first remove the contribution of $q_i(\mathbf{x})$ from $q(\mathbf{x})$:
	\begin{equation}
		q_{\backslash i}(\mathbf{x}) = \frac{q(\mathbf{x})}{q_i(\mathbf{x})},
	\end{equation}
	Then, we multiply $q_{\backslash i}(\mathbf{x})$ and $f_i(\mathbf{x})$ as
	\begin{equation}
		\hat{q}_i(\mathbf{x}) = f_i(\mathbf{x})q_{\backslash i}(\mathbf{x}) = f_i(\mathbf{x})\frac{q(\mathbf{x})}{q_i(\mathbf{x})}.
	\end{equation}
	for all $i$. Finally, we update the parameters of the approximation factor $q_i(\mathbf{x})$ such that the moments of $q(\mathbf{x})$ and $\hat{q}_i(\mathbf{x})$ are matched as
	\begin{align}
		E_{q(\mathbf{x})}[\mathbf{x}] &= E_{\hat{q}_i(\mathbf{x})}[\mathbf{x}], \\
		\mathrm{Var}_{q(\mathbf{x})}[\mathbf{x}] &= \mathrm{Var}_{\hat{q}_i(\mathbf{x})}[\mathbf{x}],
	\end{align}
	for all $i$. The sequential EP algorithm is performed until a convergence criterion is satisfied or the maximum number of iterations is reached. 
	
	In this paper, we approximate the target posterior distribution $f(\mathbf{g}) = p(\mathbf{y}_p|\mathbf{g})p(\mathbf{g})$ by a Gaussian distribution $q(\mathbf{g}) = \mathcal{CN}(\mathbf{g}|\tilde{\mathbf{m}},\tilde{\mathbf{V}})$ optimized by the EP algorithm. Then, based on the approximate distribution, AUD and CE are jointly performed and then the data is detected.
	
	\subsection{Form of the Approximation}
	In this subsection, we describe an approximate form of the target posterior distribution $f(\mathbf{g}) = f_1(\mathbf{g})f_2(\mathbf{g}) = p(\mathbf{y}_p|\mathbf{g})p(\mathbf{g})$. First, we approximate each term in $f(\mathbf{g})$ by a simple complex-Gaussian as 
	\begin{align}
		f_1(\mathbf{g}) = p(\mathbf{y}_p|\mathbf{g}) &\approx q_1(\mathbf{g}) =  \mathcal{CN}(\mathbf{g}|\tilde{\mathbf{m}}_1,\tilde{\mathbf{V}}_1),\\
		f_2(\mathbf{g}) = p(\mathbf{g}) &\approx  q_2(\mathbf{g}) = \mathcal{CN}(\mathbf{g}|\tilde{\mathbf{m}}_2,\tilde{\mathbf{V}}_2).
	\end{align} 
	
	Then, we construct an unnormalized global Gaussian approximation as 
	\begin{align}
		f(\mathbf{g}) \approx q(\mathbf{g}) = q_1(\mathbf{g})q_2(\mathbf{g})  = \mathcal{CN}(\mathbf{g}|\tilde{\mathbf{m}},\tilde{\mathbf{V}}) 
		\label{eqn:glabal_approx},
	\end{align}
	where the mean vector $\tilde{\mathbf{m}}$ and the covariance matrix 
	$\tilde{\mathbf{V}}$ are given by
	\begin{align}
		\tilde{\mathbf{V}}&=\left(\tilde{\mathbf{V}}_1^{-1}+\tilde{\mathbf{V}}_2^{-1}\right)^{-1},
		 \label{eqn:joint_param_V}\\ 
		\tilde{\mathbf{m}}&=\tilde{\mathbf{V}}\left(\tilde{\mathbf{V}}_1^{-1}\tilde{\mathbf{m}}_1+\tilde{\mathbf{V}}_2^{-1}\tilde{\mathbf{m}}_2\right).
		 \label{eqn:joint_param_m}
	\end{align}
	
	The first approximation term $q_1(\mathbf{g})$ corresponds to the Gaussian noise likelihood $p(\mathbf{y}_p|\mathbf{g})=\mathcal{CN}(\mathbf{y}_p|\mathbf{\Phi g},\sigma_w^2\mathbf{I})$. In this case, $\tilde{\mathbf{m}}_1$ and $\tilde{\mathbf{V}}_1$ can be simply characterized by the following relations
	\begin{align}
		\tilde{\mathbf{m}}_1 &= (\mathbf{\Phi}^H\mathbf{\Phi})^{-1}\mathbf{\Phi}^H\mathbf{y}_p, \label{eqn:param_V_1}\\
		\tilde{\mathbf{V}}_1 &= \sigma_w^2(\mathbf{\Phi}^H\mathbf{\Phi})^{-1}. \label{eqn:param_m_1}
	\end{align}	
	
	Using \eqref{eqn:param_V_1} and \eqref{eqn:param_m_1}, the global approximation in \eqref{eqn:glabal_approx} is rewritten as
	\begin{align}
		q(\mathbf{g}) &= q_1(\mathbf{g})q_2(\mathbf{g}) 		=\mathcal{CN}(\mathbf{g}|\tilde{\mathbf{m}},\tilde{\mathbf{V}}).
	\end{align}
	where 
	\begin{align}
		\tilde{\mathbf{V}}&=\left(\sigma_w^{-2}\mathbf{\Phi}^H\mathbf{\Phi}  +\tilde{\mathbf{V}}_2^{-1}\right)^{-1}, \label{eqn:joint_param_V_rev} \\ 
		\tilde{\mathbf{m}}&=\tilde{\mathbf{V}}\left(\sigma_w^{-2}\mathbf{\Phi}^H\mathbf{y}_p+\tilde{\mathbf{V}}_2^{-1}\tilde{\mathbf{m}}_2\right)
		 \label{eqn:joint_param_m_rev}.
	\end{align}
	
	\subsection{Iterative EP Update Rules}
\begin{figure}
		\centering
		\includegraphics[width=\linewidth]{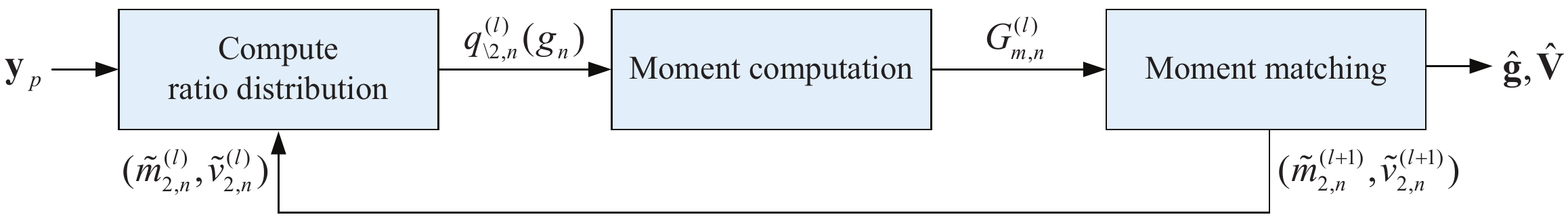}
		\caption{Flow chart of an EP iteration.}
		\label{fig:Iterative_EP}
	\end{figure}

	In this subsection, we explain the iterative EP update rules for the parameter estimations $\tilde{\mathbf{m}}_2 = [\tilde{m}_{2,1},\dots,\tilde{m}_{2,N}]^T$ and $\tilde{\mathbf{V}}_2 = \mathrm{diag}(\tilde{v}_{2,1},\dots,\tilde{v}_{2,N})$. In the first iteration, we initialize $\tilde{\mathbf{m}}_2=\mathbf{0}$ and $\tilde{\mathbf{V}}_2=\mathrm{diag}(p_1\alpha_1,\dots,p_N\alpha_N)$. In the general, say $l$-th, iteration, the pairs $(\tilde{m}_{2,n}^{({l+1})},\tilde{v}_{2,n}^{(l+1)})$ for all $n=1,\dots,N$ are updated as follows (see Fig.~\ref{fig:Iterative_EP}).
	
	First, we compute the ratio distribution $q_{\backslash 2,n}^{(l)}(g_n)$ by removing the contribution of $q_{2,n}^{(l)}(g_n)$ from $q_{n}^{(l)}(g_n) = \mathcal{CN}(g_n|\tilde{m}_{n}^{({l})},\tilde{V}_{nn}^{(l)})$, which corresponds to the $n$-th marginal of $q^{(l)}(\mathbf{g})$. That is,
	\begin{align}
		q_{\backslash 2,n}^{(l)}(g_n) 
		= \frac{q_{n}^{(l)}(g_n)}{q_{2,n}^{(l)}(g_n)}
		=\frac{\mathcal{CN}(g_n|\tilde{m}_{n}^{({l})},\tilde{V}_{nn}^{(l)})}{\mathcal{CN}(g_n|\tilde{m}_{2,n}^{({l})},\tilde{v}_{2,n}^{(l)})}
		=\mathcal{CN}(g_n|\tilde{m}_{\backslash 2,n}^{(l)},\tilde{v}_{\backslash 2,n}^{(l)}),
	\end{align}
	where the mean and variance of $q_{\backslash 2,n}^{(l)}(g_n)$ are given by
	\begin{align}
		\tilde{v}_{\backslash 2,n}^{(l)} &= \left[(\tilde{V}_{nn}^{(l)})^{-1} -  (\tilde{v}_{2,n}^{(l)})^{-1}\right]^{-1}, \label{eqn:var_itr_update}\\
		\tilde{m}_{\backslash 2,n}^{(l)} &= \tilde{v}_{\backslash  2,n}^{(l)}\left[(\tilde{V}_{nn}^{(l)})^{-1}\tilde{m}_{n}^{(l)} -  (\tilde{v}_{2,n}^{(l)})^{-1}\tilde{m}_{2,n}^{({l})}\right].
	\end{align}
	
	Second, we update $(\tilde{m}_{2,n}^{({l+1})},\tilde{v}_{2,n}^{(l+1)})$ to match the mean and variance of $q_{n}^{(l+1)}(g_n)$ to those of the following distribution
	\begin{equation}
		\hat{q}_{n}^{(l)}(g_n) = f_{2,n}(g_n)q_{\backslash 2,n}^{(l)}(g_n).
	\end{equation}
	To compute the mean $E_{q}^{(l)}[g_n]$ and variance $V_{q}^{(l)}[g_n]$ of the distribution $\hat{q}_{n}^{(l)}(g_n)$, we firstly compute $G_{m,n}^{(l)}$, the $m$-th ($m=0,1,2$) moments of $\hat{q}_{n}^{(l)}(g_n)$ with respect to $g_n$ (see Appendix A for details)
	\begin{align}
		G_{0,n}^{(l)} &= \int_{-\infty}^{\infty} \hat{q}_{n}^{(l)}(g_n)\mathrm{d}g_n \nonumber\\
		&=(1-p_{n})\mathcal{CN}(0|\tilde{m}_{\backslash 2,n}^{(l)},\tilde{v}_{\backslash 2,n}^{(l)})+p_{n}\mathcal{CN}(0|\tilde{m}_{\backslash 2,n}^{(l)},\alpha_n+\tilde{v}_{\backslash 2,n}^{(l)}),\\
		G_{1,n}^{(l)} &= \int_{-\infty}^{\infty} g_{n} \hat{q}_{n}^{(l)}(g_n) \mathrm{d}g_n \nonumber\\
		&=p_{n}\mathcal{CN}(0|\tilde{m}_{\backslash 2,n}^{(l)},\alpha_n+\tilde{v}_{\backslash 2,n}^{(l)})
		\frac{\tilde{m}_{\backslash 2,n}^{(l)}\alpha_n}{\alpha_n+\tilde{v}_{\backslash 2,n}^{(l)}},\\
		G_{2,n}^{(l)} &= \int_{-\infty}^{\infty} \left|g_{n}\right|^2 \hat{q}_{n}^{(l)}(g_n)\mathrm{d}g_n \nonumber\\
		&=p_{n}\mathcal{CN}(0|\tilde{m}_{\backslash 2,n}^{(l)},\alpha_n+\tilde{v}_{\backslash 2,n}^{(l)})\cdot\left(\left|\frac{\tilde{m}_{\backslash 2,n}^{(l)}\alpha_n}{\alpha_n+\tilde{v}_{\backslash  2,n}^{(l)}}\right|^2+\frac{\alpha_n\tilde{v}_{\backslash 2,n}^{(l)}}{\alpha_n+\tilde{v}_{\backslash 2,n}^{(l)}}\right).
	\end{align}
	Then, $E_{q}^{(l)}[g_n]$ and $V_{q}^{(l)}[g_n]$ are calculated as
	\begin{align}
		E_{q}^{(l)}[g_n] &= \frac{G_{1,n}^{(l)}}{G_{0,n}^{(l)}}, \label{eqn:E_calc}\\
		V_{q}^{(l)}[g_n] &= \frac{G_{2,n}^{(l)}}{G_{0,n}^{(l)}}-\left|E_{q}^{(l)}[g_n]\right|^2. \label{eqn:V_calc}
	\end{align}
	
	Finally, we update the parameter pair $(\tilde{m}_{2,n}^{({l+1})},\tilde{v}_{2,n}^{(l+1)})$ such that the unnormalized distribution $q_{n}^{(l+1)}(g_n) = q_{2,n}^{(l+1)}(g_n)q_{\backslash 2,n}^{(l)}(g_n)$ has the mean $E_{q}^{(l)}[g_n]$ and variance $V_{q}^{(l)}[g_n]$ obtained from \eqref{eqn:E_calc} and \eqref{eqn:V_calc}, respectively. The corresponding solution is
	\begin{align}
		\tilde{v}_{2,n}^{(l+1)} &= \left[V_{q}^{(l)}[g_n]^{-1} - (\tilde{v}_{\backslash 2,n}^{(l)})^{-1}\right]^{-1}, \label{eqn:var_update}\\
		\tilde{m}_{2,n}^{({l+1})} &= \tilde{v}_{2,n}^{(l+1)}\left[V_{q}^{(l)}[g_n]^{-1}E_{q}^{(l)}[g_n] - (\tilde{v}_{\backslash 2,n}^{(l)})^{-1}\tilde{m}_{\backslash 2,n}^{(l)}\right]. \label{eqn:mean_update}
	\end{align}
	
	We stop the algorithm when the variation of the mean vector $\tilde{\mathbf{m}}$ is smaller than the given threshold $\epsilon$ (e.g., $\epsilon = 10^{-4}$) or the maximum number of iterations has been reached. For the robust convergence of the proposed algorithm, we can smooth the parameter update by taking a combination of the previous and new parameter values as
	\begin{align}
		\tilde{m}_{2,n}^{(l+1)} &= \beta \tilde{m}_{2,n}^{\textrm{new}} + (1-\beta) \tilde{m}_{2,n}^{(l)},\\
		\tilde{v}_{2,n}^{(l+1)} &= \beta \tilde{v}_{2,n}^{\textrm{new}} + (1-\beta) \tilde{v}_{2,n}^{(l)},
	\end{align}
	where $\beta \in [0,1]$ is the smoothing parameter and $(\tilde{m}_{2,n}^{\textrm{new}},\tilde{v}_{2,n}^{\textrm{new}})$ is the new parameter pair computed in \eqref{eqn:var_update} and \eqref{eqn:mean_update}.
	
	After the convergence of the EP algorithm, an approximate distribution of $f(\mathbf{g})=p(\mathbf{y}_p|\mathbf{g})p(\mathbf{g})$ is obtained as
	\begin{align}
		q(\mathbf{g}) &=\mathcal{CN}(\mathbf{g}|\tilde{\mathbf{m}},\tilde{\mathbf{V}}),
	\end{align}
	where $\tilde{\mathbf{m}}$ and $\tilde{\mathbf{V}}$ are obtained from \eqref{eqn:joint_param_V_rev} and \eqref{eqn:joint_param_m_rev}. The final solution of the iterative EP estimator $\hat{\mathbf{g}}$ is the mean vector of $q(\mathbf{g})$ given by
	\begin{align}
		\hat{\mathbf{g}}&=E_{q(\mathbf{g})}[\mathbf{g}]=\tilde{\mathbf{m}}.
	\end{align}
	and the covariance matrix of $\hat{\mathbf{g}}$ is $\tilde{\mathbf{V}}$. Once $\hat{\mathbf{g}}$ is obtained, by performing the log-likelihood test on $\hat{\mathbf{g}}$, we detect the active devices. Then, we find the corresponding channel of each active device.
	
	\subsection{Active User Detection and Channel Estimation}
	To perform AUD, we employ the likelihood ratio test on the vector $\hat{\mathbf{g}}$. The hypothesis testing to find out the active device is given by
	\begin{align}
		\left\{ \begin{array}{ll}
		H_1 &: a_n = 1,\quad\textrm{active device},\\
		H_0 &: a_n = 0,\quad\textrm{inactive device},
		\end{array} \right.
	\end{align}
	and the corresponding log-likelihood ratio test is
	\begin{align}
		\textrm{LLR}(\hat{g}_n) = \log\left(\frac{p_{\hat{g}_n|a_n}(\hat{g}_n|a_n \neq 0)}{p_{\hat{g}_n|a_n}(\hat{g}_n|a_n = 0)}\right) \underset{H_0}{\overset{H_1}{\gtrless}} 0.
		\label{eqn:LLR-test}
	\end{align}
	
	Since the $n$-th component $\hat{g}_n$ of $\hat{\mathbf{g}}$ has the variance $\tilde{V}_{nn}$ (the $n$-th diagonal of $\tilde{\mathbf{V}}$), the likelihood probabilities of $\hat{g}_n$ given $a_n \neq 0$ or $a_n = 0$ are given, respectively, by
	\begin{align}
		p_{\hat{g}_n|a_n}(\hat{g}_n|a_n \neq 0) &= \frac{1}{\pi(\alpha_n+\tilde{V}_{nn})}\exp\left(\frac{-\left|\hat{g}_n\right|^2}{\alpha_n+\tilde{V}_{nn}}\right),\\
		p_{\hat{g}_n|a_n}(\hat{g}_n|a_n = 0) &= \frac{1}{\pi\tilde{V}_{nn}}\exp\left(\frac{-\left|\hat{g}_n\right|^2}{\tilde{V}_{nn}}\right).
	\end{align}
	and the corresponding log-likelihood ratio is
	\begin{align}
		\textrm{LLR}(\hat{g}_n) &=\log\left(\frac{p_{\hat{g}_n|a_n}(\hat{g}_n|a_n \neq 0)}{p_{\hat{g}_n|a_n}(\hat{g}_n|a_n = 0)}\right), \nonumber\\ 
		&=\log\left(\frac{\tilde{V}_{nn}}{\alpha_n+\tilde{V}_{nn}}\exp\left(\left|\hat{g}_n\right|^2\left(\frac{1}{\tilde{V}_{nn}}-\frac{1}{\alpha_n+\tilde{V}_{nn}}\right)\right)\right) \underset{H_0}{\overset{H_1}{\gtrless}} 0. 
		\label{eqn:LLR-test-2}
	\end{align} 
	
	The log-likelihood ratio test in \eqref{eqn:LLR-test-2} can be simplified to
	\begin{align}
		\left|\hat{g}_n\right|^2 \underset{H_0}{\overset{H_1}{\gtrless}} \theta_n,
	\end{align}
	where
	\begin{align}
		\theta_n = \log(1+\frac{\alpha_n}{\tilde{V}_{nn}})/(\frac{1}{\tilde{V}_{nn}}-\frac{1}{\alpha_n+\tilde{V}_{nn}}).
		\label{eqn:AUD_threshold}
	\end{align}
	
	The estimate of activity vector $\hat{\mathbf{a}}$ is obtained after the thresholding of each element of $\hat{\mathbf{g}}$ as
	\begin{align}
		\hat{a}_n &= \left\{ \begin{array}{ll}
		1 & \textrm{if $|\hat{g}_n|^2 \geq \theta_n$},\\
		0 & \textrm{if $|\hat{g}_n|^2 < \theta_n$}.\\
	\end{array} \right.
	\end{align}
	If the $n$-th device is detected to be active, we use $\hat{g}_n$ as the CSI estimate of the device.
	
	\subsection{Data Detection}
	Based on the knowledge of user activities and channels obtained in the AUD/CE phase, data symbols of active devices are detected. The measurement signal $\mathbf{y}_{d}^{[i]}$ of the $i$-th data symbol vector is given by
	\begin{align}
		\mathbf{y}_{d}^{[i]}=\sum_{n \in \mathcal{N}}\mathbf{s}_n h_n x_{d,n}^{[i]} + \mathbf{w}_{d}^{[i]}, \quad i=1,\dots,J,
		\label{eqn:y_d}
	\end{align}	
	where $\mathcal{N}=\{n_1,\dots,n_{|\mathcal{N}|}\}$ is the true support\footnote{For example, if $\mathbf{a} = [1\:0\:1\:0\:0\:0]^T$, then the support $T$ is $T=\{1,3\}$.} (index set of active devices) of active devices and $x_{d,n}^{[i]}$ is the $i$-th data symbol of the $n$-th device drawn from the finite alphabet $ \mathcal{A}$, respectively.
	
	Let $\mathbf{S}_{\mathcal{N}} = [ \mathbf{s}_{n_1} ,\dots,\mathbf{s}_{n_{|\mathcal{N}|}} ]$ be the matrix containing the spreading sequences of active devices and $\mathbf{h}_{\mathcal{N}} = [ h_{n_1} ,\dots,h_{n_{|\mathcal{N}|}} ]^T$ be the vector containing the channels of active devices. Then \eqref{eqn:y_d} can be rewritten as 
	\begin{align}
		\mathbf{y}_{d}^{[i]}=\mathbf{S}_{\mathcal{N}} \textrm{diag}(\mathbf{h}_{\mathcal{N}})\mathbf{x}_{d,\mathcal{N}}^{[i]} + \mathbf{w}_d^{[i]}
	\end{align}	
	where $\mathbf{x}_{d,\mathcal{N}}^{[i]} \in \mathcal{A}^{|\mathcal{N}|}$ is the $i$-th data symbols of all active devices.
	
	Let $\hat{\mathcal{N}}=\{\hat{n}_1,\dots,\hat{n}_{|\hat{\mathcal{N}}|}\}$ be the estimated support. Then let $\mathbf{S}_{\hat{\mathcal{N}}}$ be the spreading sequences of devices in $\hat{\mathcal{N}}$ and $\hat{\mathbf{h}}_{\hat{\mathcal{N}}}$ be the estimated channels of devices in $\hat{\mathcal{N}}$. Further, let $\tilde{\mathbf{L}}_{\hat{\mathcal{N}}}$ be
	\begin{align}
		\tilde{\mathbf{L}}_{\hat{\mathcal{N}}} = \mathbf{S}_{\hat{\mathcal{N}}} \textrm{diag}(\hat{\mathbf{h}}_{\hat{\mathcal{N}}}) = (\mathbf{s}_{\hat{n}_1}\hat{h}_{\hat{n}_1},\dots,\mathbf{s}_{\hat{n}_{|\hat{\mathcal{N}}|}}\hat{h}_{\hat{n}_{|\hat{\mathcal{N}}|}}),
	\end{align}
	then the BS employs the linear MMSE detector followed by the quantization to detect the data symbols as
	\begin{align}
		\hat{\mathbf{x}}_{d,\hat{\mathcal{N}}}^{[i]} &= \mathnormal{Q}_{\mathcal{A}}\left(\left(\tilde{\mathbf{L}}_{\hat{\mathcal{N}}}^H\tilde{\mathbf{L}}_{\hat{\mathcal{N}}} + \sigma_w^2\mathbf{I}_{|\mathcal{N}|}\right)^{-1}\tilde{\mathbf{L}}_{\hat{\mathcal{N}}}^H\mathbf{y}_{d}^{[i]}\right), \quad i=1,\dots,J,
	\end{align}
	where $\mathnormal{Q}_{\mathcal{A}}(\cdot)$ is the quantization operator on the finite alphabet $\mathcal{A}$. In Table \ref{table:ALG_SUMMARY}, we summarize the proposed algorithm.
	
	\begin{table}[]
		\renewcommand{\arraystretch}{1.3}
		\caption{SUMMARY OF THE PROPOSED ALGORITHM}
		\label{table:ALG_SUMMARY}
		\centering
		\begin{tabularx}{\linewidth}{l}
			\hline
			\textbf{Input:} The pilot measurements $\mathbf{y}_p$, the data measurements $\mathbf{y}_{d}^{[i]}$, the spreading sequence \{$\mathbf{s}_n$\}, the pilot symbol \{$x_{p,n}$\}, \\
			\qquad the activity probability \{$p_n$\}, the channel variances \{$\alpha_n$\}, the noise variance $\sigma_w^2$, the finite alphabet $\mathcal{A}$ \\
			\hline
			
			\textbf{Output:} The estimated support of active devices $\hat{\mathcal{N}}$, the estimated channel vector $\hat{\mathbf{h}}_{\hat{\mathcal{N}}}$, the detected data symbol vectors $\hat{\mathbf{x}}_{d,\hat{\mathcal{N}}}^{[i]}$ \\
			\hline
			
			\textbf{Subscript and superscript:} The device index $n$, the iteration index $l$, the data symbol index $i$\\
			\hline
			
			\textbf{Step 1: (Initialization)} \\
			\qquad $\mathbf{\Phi} = 
			[\mathbf{s}_1 x_{p,1},\dots,\mathbf{s}_N x_{p,N}]$\\
			\qquad $\tilde{\mathbf{m}}_{2}^{(1)}=\mathbf{0}_N$,	$\tilde{\mathbf{V}}_{2}^{(1)}=\mathrm{diag}(p_1\alpha_1,\dots,p_N\alpha_N)$ \\
			\qquad 
			$\tilde{\mathbf{V}}^{(1)}=\left(\sigma_w^{-2}\mathbf{\Phi}^H\mathbf{\Phi}+\tilde{(\mathbf{V}}_2^{(1)})^{-1}\right)^{-1}$,  
			$\tilde{\mathbf{m}}^{(1)}=\tilde{\mathbf{V}}\left(\sigma_w^{-2}\mathbf{\Phi}^H\mathbf{y}_p+\tilde{(\mathbf{V}}_2^{(1)})^{-1}\tilde{\mathbf{m}}_2\right)$
			 \\
			
			\textbf{Step 2: (Compute ratio distribution)} Compute the parameters of the distribution $q_{\backslash 2,n}^{(l)}$. \\
			\qquad $\tilde{v}_{\backslash 2,n}^{(l)} = \left[(\tilde{V}_{nn}^{(l)})^{-1} - (\tilde{v}_{2,n}^{(l)})^{-1}\right]^{-1}$\\ 
			\qquad $\tilde{m}_{\backslash 2,n}^{(l)} = \tilde{v}_{\backslash 2,n}^{(l)}\left[(\tilde{V}_{nn}^{(l)})^{-1}\tilde{m}_{n}^{(l)} - (\tilde{v}_{2,n}^{(l)})^{-1}\tilde{m}_{2,n}^{({l})}\right]$ \\
			
			\textbf{Step 3: (Moment computation)} Compute the mean $E_{q}^{(l)}[g_n]$ and variance $V_{q}^{(l)}[g_n]$ of the distribution $f_{2,n}q_{\backslash 2,n}^{(l)}$.\\
			\qquad $G_{0,n}^{(l)} = (1-p_{n})\mathcal{CN}(0|\tilde{m}_{\backslash 2,n}^{(l)},\tilde{v}_{\backslash 2,n}^{(l)})+p_{n}\mathcal{CN}(0|\tilde{m}_{\backslash 2,n}^{(l)},\alpha_n+\tilde{v}_{\backslash 2,n}^{(l)})$ \\
			\qquad $G_{1,n}^{(l)} = p_{n}\mathcal{CN}(0|\tilde{m}_{\backslash 2,n}^{(l)},\alpha_n+\tilde{v}_{\backslash 2,n}^{(l)})\frac{\tilde{m}_{\backslash 2,n}^{(l)}\alpha_n}{\alpha_n+\tilde{v}_{\backslash 2,n}^{(l)}}$ \\
			\qquad $G_{2,n}^{(l)} = p_{n}\mathcal{CN}(0|\tilde{m}_{\backslash 2,n}^{(l)},\alpha_n+\tilde{v}_{\backslash 2,n}^{(l)})\cdot\left(\left|\frac{\tilde{m}_{\backslash 2,n}^{(l)}\alpha_n}{\alpha_n+\tilde{v}_{\backslash 2,n}^{(l)}}\right|^2+\frac{\alpha_n\tilde{v}_{\backslash 2,n}^{(l)}}{\alpha_n+\tilde{v}_{\backslash 2,n}^{(l)}}\right)$ \\
			\qquad $E_{q}^{(l)}[g_n] = \frac{G_{1,n}^{(l)}}{G_{0,n}^{(l)}}$\\ 
			\qquad $V_{q}^{(l)}[g_n] = \frac{G_{2,n}^{(l)}}{G_{0,n}^{(l)}}-\left|E_{q}^{(l)}[g_n]\right|^2$
			 \\
			
			\textbf{Step 4: (Moment matching)} Update the parameters 
			$\tilde{v}_{2,n}^{({l+1})}$ and $\tilde{m}_{2,n}^{(l+1)}$. \\
			\qquad $\tilde{v}_{2,n}^{(l+1)} = \left[V_{q}^{(l)}[g_n]^{-1} - (\tilde{v}_{\backslash 2,n}^{(l)})^{-1}\right]^{-1}$\\ 
			\qquad $\tilde{m}_{2,n}^{({l+1})} = \tilde{v}_{2,n}^{(l+1)}\left[V_{q}^{(l)}[g_n]^{-1}E_{q}^{(l)}[g_n] - (\tilde{v}_{\backslash 2,n}^{(l)})^{-1}\tilde{m}_{\backslash 2,n}^{(l)}\right]$ \\
			
			\textbf{Step 5: (Iteration)} Repeat \textbf{Step 2 - 4} until stopping criteria, $\forall n$. After the convergence of $\tilde{\mathbf{V}}_2$ and $\tilde{\mathbf{m}}_2$, we have\\
			\qquad $\tilde{\mathbf{V}}=\left(\sigma_w^{-2}\mathbf{\Phi}^H\mathbf{\Phi} + \tilde{\mathbf{V}}_2^{-1}\right)^{-1}$\\
			\qquad $\hat{\mathbf{g}}=\tilde{\mathbf{m}}=\tilde{\mathbf{V}}\left(\sigma_w^{-2}\mathbf{\Phi}^H\mathbf{y}_p+\tilde{\mathbf{V}}_2^{-1}\tilde{\mathbf{m}}_2\right)$\\
			
			\textbf{Step 6: (Active user detection channel estimation)} 
			Threshold each element of $\hat{\mathbf{g}}$ using the threshold $\theta_n$ in \eqref{eqn:AUD_threshold}.\\
			\qquad $\hat{\mathcal{N}} = \{n | |\hat{g}_n|^2 \geq \theta_n\}$ \\
			\qquad $\hat{h}_n = \hat{g}_n$, $\forall n \in \hat{\mathcal{N}}$ \\
			
			\textbf{Step 7: (Data detection)} Use the linear MMSE detector to detect the data symbols.\\
			\qquad $\tilde{\mathbf{L}}_{\hat{\mathcal{N}}} = \left[\mathbf{s}_{1}\hat{h}_{1},\dots,\mathbf{s}_{n}\hat{h}_{n},\dots,\mathbf{s}_{|\hat{\mathcal{N}}|}\hat{h}_{|\hat{\mathcal{N}}|}\right]$ for $n \in \hat{\mathcal{N}}$ \\
			\qquad $\hat{\mathbf{x}}_{d,\hat{\mathcal{N}}}^{[i]} = \mathnormal{Q}_{\mathcal{A}}\left(\left(\tilde{\mathbf{L}}_{\hat{\mathcal{N}}}^H\tilde{\mathbf{L}}_{\hat{\mathcal{N}}} + \sigma_w^2\mathbf{I}_{|\mathcal{N}|}\right)^{-1}\tilde{\mathbf{L}}_{\hat{\mathcal{N}}}^H\mathbf{y}_{d}^{[i]}\right)$ for $i \in [1:J]$\\
			\hline
		\end{tabularx}
	\end{table}

	\subsection{Comments on Complexity}\label{sec:complexity}
	In this subsection, we briefly discuss the computational complexity of the proposed algorithm. In each EP iteration, computations of marginals and moments, and updating the parameter pairs $(\tilde{m}_{2,n}^{({l})},\tilde{v}_{2,n}^{(l)})$ for all $n=1,\dots,N$ has a small linear complexity of $\mathcal{O}(N)$ because they are composed only of arithmetic operations such as addition, subtraction, multiplication, and division. The complexity of the EP algorithm is dominated by the computation of the covariance matrix $\tilde{\mathbf{V}}$ in \eqref{eqn:joint_param_V_rev} and the mean vector $\tilde{\mathbf{m}}$ in \eqref{eqn:joint_param_m_rev}. When $N$ is large, direct computation of $\tilde{\mathbf{V}}$ and $\tilde{\mathbf{m}}$ is computationally burdensome due to the computationally expensive matrix inversion. When $M < N$ and $\tilde{\mathbf{V}}_2$ is diagonal, the Woodbury matrix identity \cite{Hager89} offers an efficient way to compute $\tilde{\mathbf{V}}$. To be specific,
	\begin{align}
		\tilde{\mathbf{V}} 
		&=\left(\sigma_w^{-2}\mathbf{\Phi}^H\mathbf{\Phi}  +\tilde{\mathbf{V}}_2^{-1}\right)^{-1} \nonumber\\
		&=\tilde{\mathbf{V}}_2-\tilde{\mathbf{V}}_2\mathbf{\Phi}^H\left(\sigma_w^2\mathbf{I}_M+\mathbf{\Phi}\tilde{\mathbf{V}}_2\mathbf{\Phi}^H\right)^{-1}\mathbf{\Phi}\tilde{\mathbf{V}}_2.\label{eqn:woodbury}
	\end{align}
	Note that $\left(\sigma_w^2\mathbf{I}_M+\mathbf{\Phi}\tilde{\mathbf{V}}_2\mathbf{\Phi}^H\right)^{-1}$ can be computed in the order of $\mathcal{O}(M^3)$ by a Cholesky decomposition. Thus, the computational complexity order to compute \eqref{eqn:woodbury} is reduced from $\mathcal{O}(N^3)$ to $\mathcal{O}(MN^2)$. 
	
	Similarly, the posterior mean $\tilde{\mathbf{m}}$ can be computed in the order of $\mathcal{O}(MN^2)$ as
	\begin{align}
		\tilde{\mathbf{m}}&=\tilde{\mathbf{V}}\left(\sigma_w^{-2}\mathbf{\Phi}^H\mathbf{y}+\tilde{\mathbf{V}}_2^{-1}\tilde{\mathbf{m}}_2\right) \nonumber\\
		&=\left[\tilde{\mathbf{V}}_2-\tilde{\mathbf{V}}_2\mathbf{\Phi}^H\left(\sigma_w^2\mathbf{I}_M+\mathbf{\Phi}\tilde{\mathbf{V}}_2\mathbf{\Phi}^H\right)^{-1}\mathbf{\Phi}\tilde{\mathbf{V}}_2\right]\left(\sigma_w^{-2}\mathbf{\Phi}^H\mathbf{y}+\tilde{\mathbf{V}}_2^{-1}\tilde{\mathbf{m}}_2\right)\nonumber\\
		&=\tilde{\mathbf{V}}_2\bm{\eta}-\tilde{\mathbf{V}}_2\mathbf{\Phi}^H\left(\sigma_w^2\mathbf{I}_M+\mathbf{\Phi}\tilde{\mathbf{V}}_2\mathbf{\Phi}^H\right)^{-1}\mathbf{\Phi}\tilde{\mathbf{V}}_2\bm{\eta},
	\end{align}
	where $\bm{\eta} =\sigma_w^{-2}\mathbf{\Phi}^H\mathbf{y}+\tilde{\mathbf{V}}_2^{-1}\tilde{\mathbf{m}}_2$. When the EP algorithm runs $L$ iterations ($L=2\sim3$), the total computational complexity order of the EP approximation would be $\mathcal{O}(LMN^2)$.
	
	\section{Simulation Results and Discussions}
	\subsection{Simulation Setup}
	In our simulations, we simulate underdetermined mMTC systems with $N = 128$ MTDs and $M$-dimensional unit-norm random spreading sequences ($M < N$). Devices are randomly located in a cell with a radius 200\textrm{m}. The pathloss component of the wireless channel between the $n$-th device and the BS is modeled as $\alpha_n = -128.1 - 36.7\log_{10}(d_n)$ in \textrm{dB} scale where $d_n$ is the distance (in \textrm{km}) between the $n$-th device and the BS. The noise spectral density and transmission bandwidth are set to $-170$ \textrm{dBm/Hz} and $1$ \textrm{MHz}, respectively. The activity probabilities of all devices are set to $p_a$ (i.e., $p_n = p_a$ for all n). Note that the proposed technique can also be applied to scenarios with heterogeneous activity probabilities. The number of data symbols in a frame is set to $J = 9$. Data symbols of active devices are modulated with quadrature phase shift keying (QPSK) and the pilot symbol $x_{p,n}$ is set to 1 for simplicity. 
	
	As performance measures, we consider the activity error rate (AER), the net normalized mean squared error (NNMSE), and the net symbol error rate (NSER). The AER refers to the percentage of errors (both missed detections and false alarms) in AUD. The NNMSE, defined as the NMSE of the estimated channels of active devices, is computed as $\frac{\|\hat{\mathbf{h}}_{\mathcal{N}} - \mathbf{h}_{\mathcal{N}}\|_2^2}{\|\mathbf{h}_{\mathcal{N}}\|_2^2}$ where $\mathcal{N}$ is the support of active devices. The NSER refers to the data symbol error rate of active devices.
	
	As a reference, we compare the proposed EP estimator with the orthogonal matching pursuit (OMP), and approximate message passing (AMP). As Bayesian sparse recovery algorithms exploiting the prior distribution of user activities and channels, we also test the sparse Bayesian learning (SBL)\cite{Tipping01,Wipf04} and Bayesian compressive sensing (BCS)\cite{Ji08,Ji09}. SBL implicitly estimates the prior distribution from the received signal and computes a MAP estimate of the target signal by expectation maximization (EM) iterations. BCS is an extension of SBL accounting for the contribution of the noise variance. Lastly, as the best achievable bound of the estimation techniques, we use the Oracle minimum mean squared error (MMSE) detector. Since support information is given in the Oracle MMSE detector, it can solve the problem in the overdetermined setup. Each point of the performance figure represents an average of at least 100,000 realizations of the activities and channels.
	
	\begin{figure}[t]
		\centering
		\subfloat[]{\includegraphics[width=0.8\linewidth]{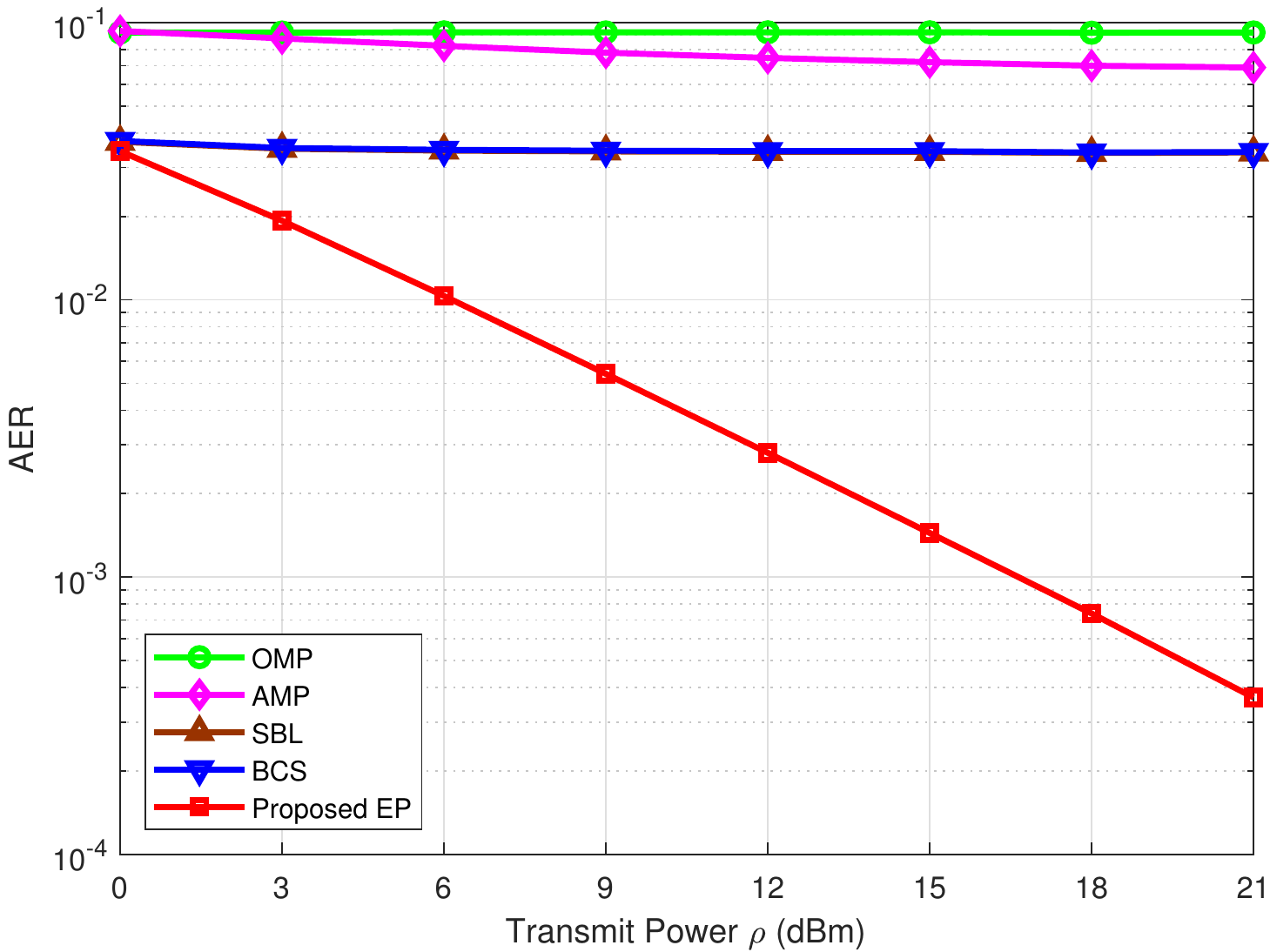}}\label{fig:sim_1_a}
		\vfill
		\subfloat[]{\includegraphics[width=0.8\linewidth]{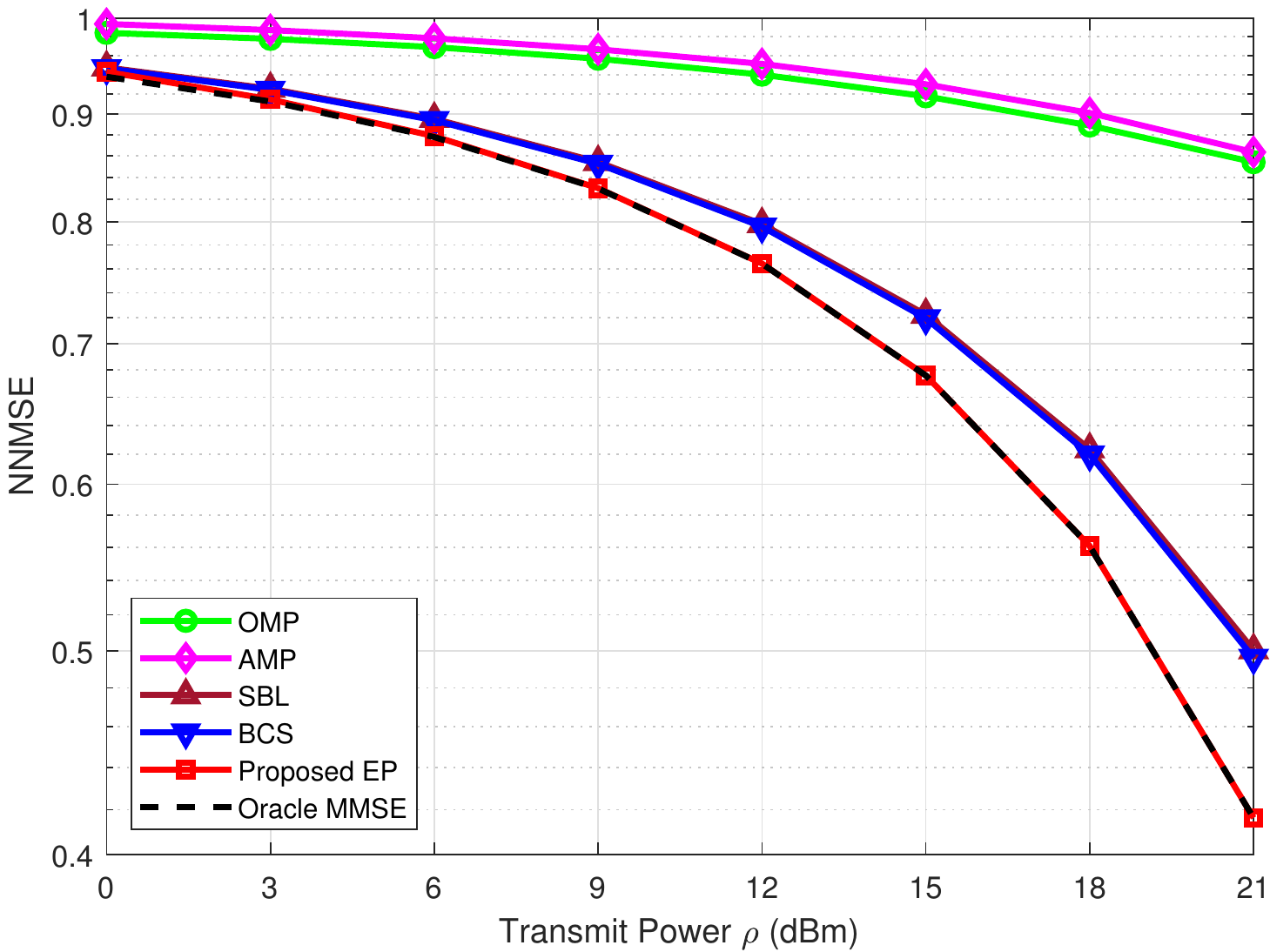}}
		\label{fig:sim_1_b}
	\end{figure}
	\begin{figure}[t]\ContinuedFloat
		\centering
		\subfloat[]{\includegraphics[width=0.8\linewidth]{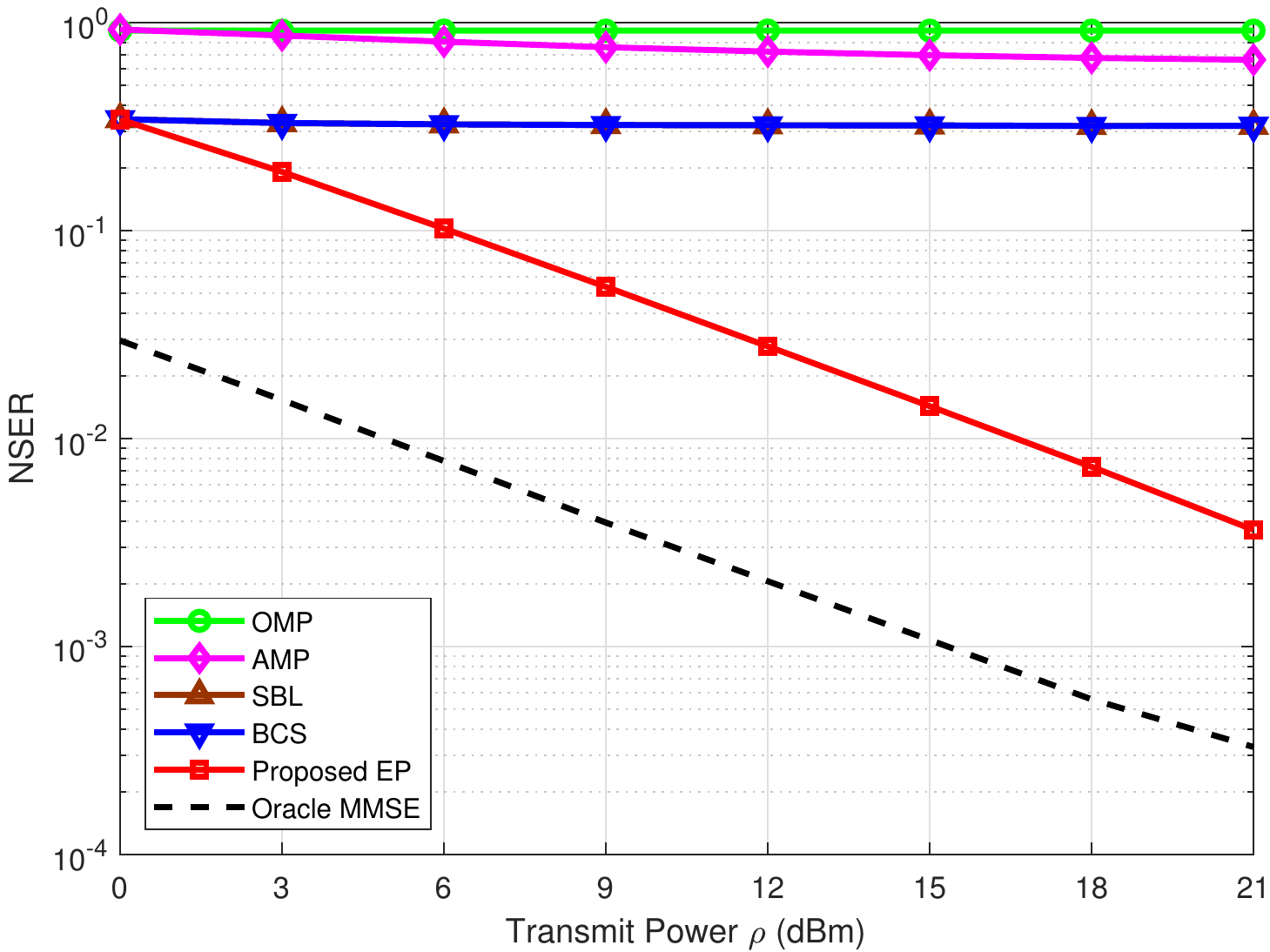}}\label{fig:sim_1_c}
		\addtocounter{figure}{+1}
		\caption{(a) AER, (b) NNMSE, and (c) NSER as a function of the transmit power 
			$\rho$.}
		\label{fig:sim_1}
	\end{figure}
	
	\begin{figure}[t]
		\centering
		\subfloat[]{\includegraphics[width=0.8\linewidth]{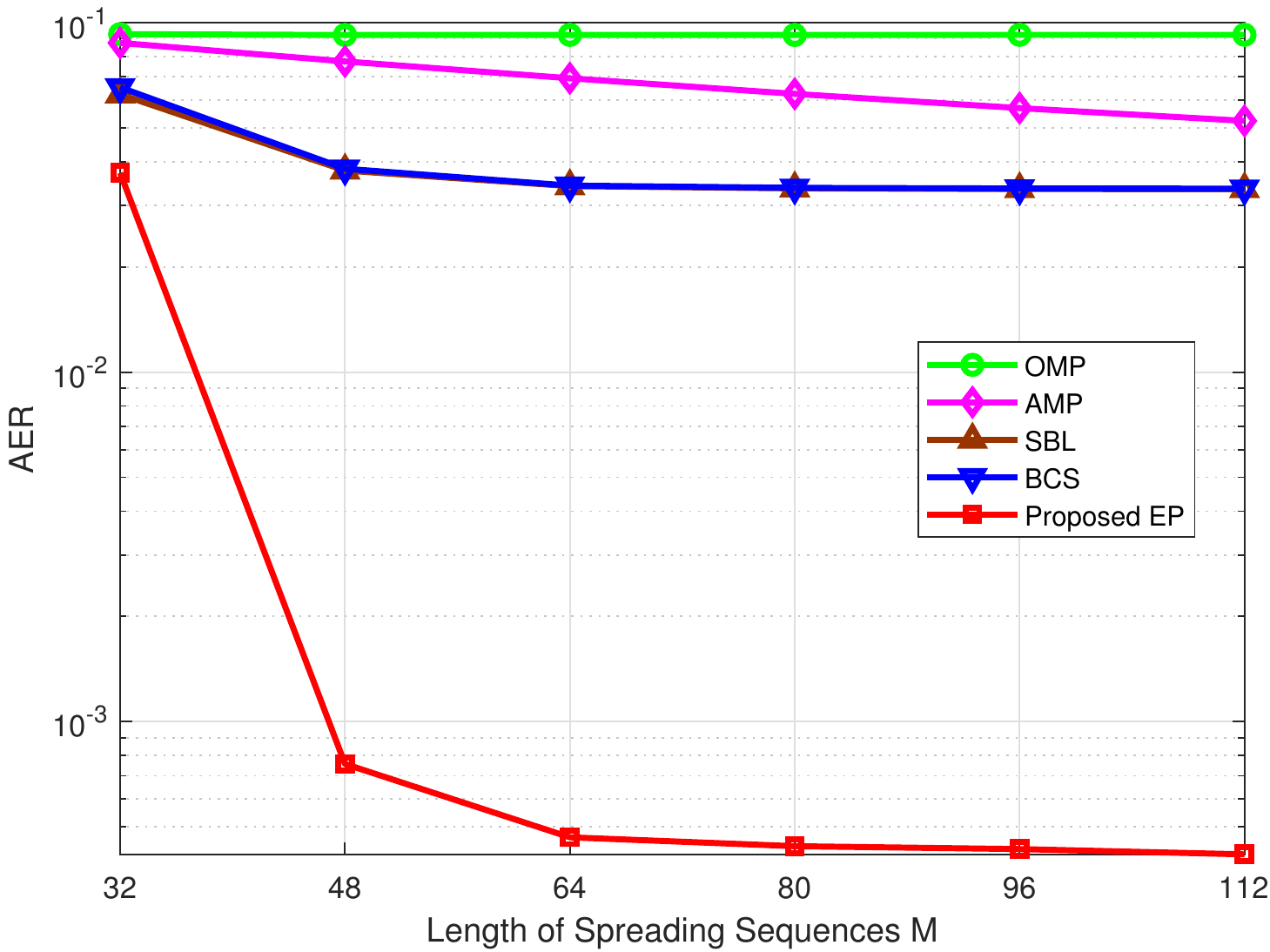}}\label{fig:sim_2_a}
		\vfill
		\subfloat[]{\includegraphics[width=0.8\linewidth]{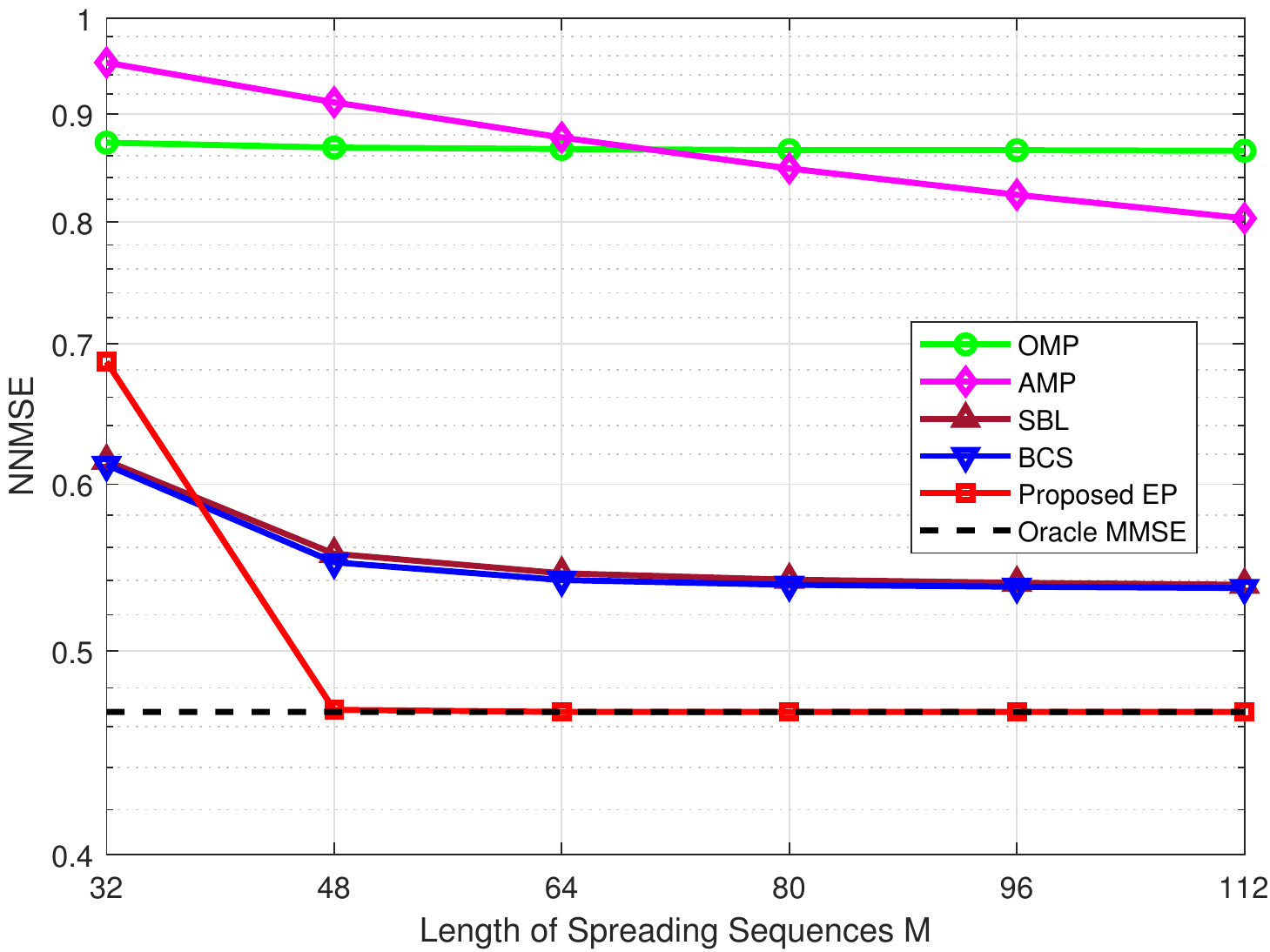}}\label{fig:sim_2_b}
	\end{figure}
	\begin{figure}[t]\ContinuedFloat
		\centering
		\subfloat[]{\includegraphics[width=0.8\linewidth]{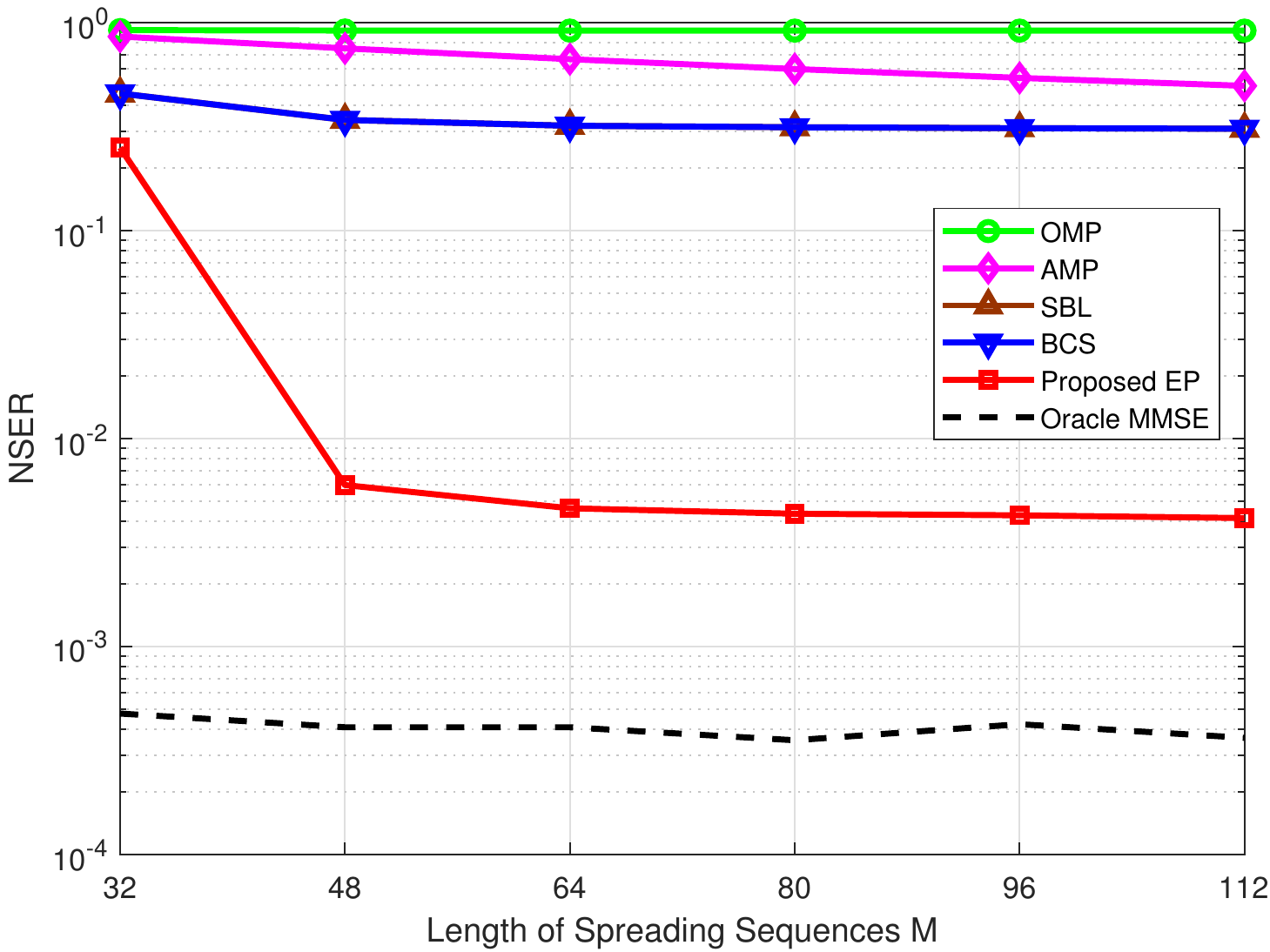}}
		\label{fig:sim_2_c}
		\addtocounter{figure}{+1}
		\caption{(a) AER, (b) NNMSE, and (c) NSER as a function of the length of 
			spreading sequences $M$.}
		\label{fig:sim_2}
	\end{figure}
	
	\begin{figure}[t]
		\centering
		\subfloat[]{\includegraphics[width=0.8\linewidth]{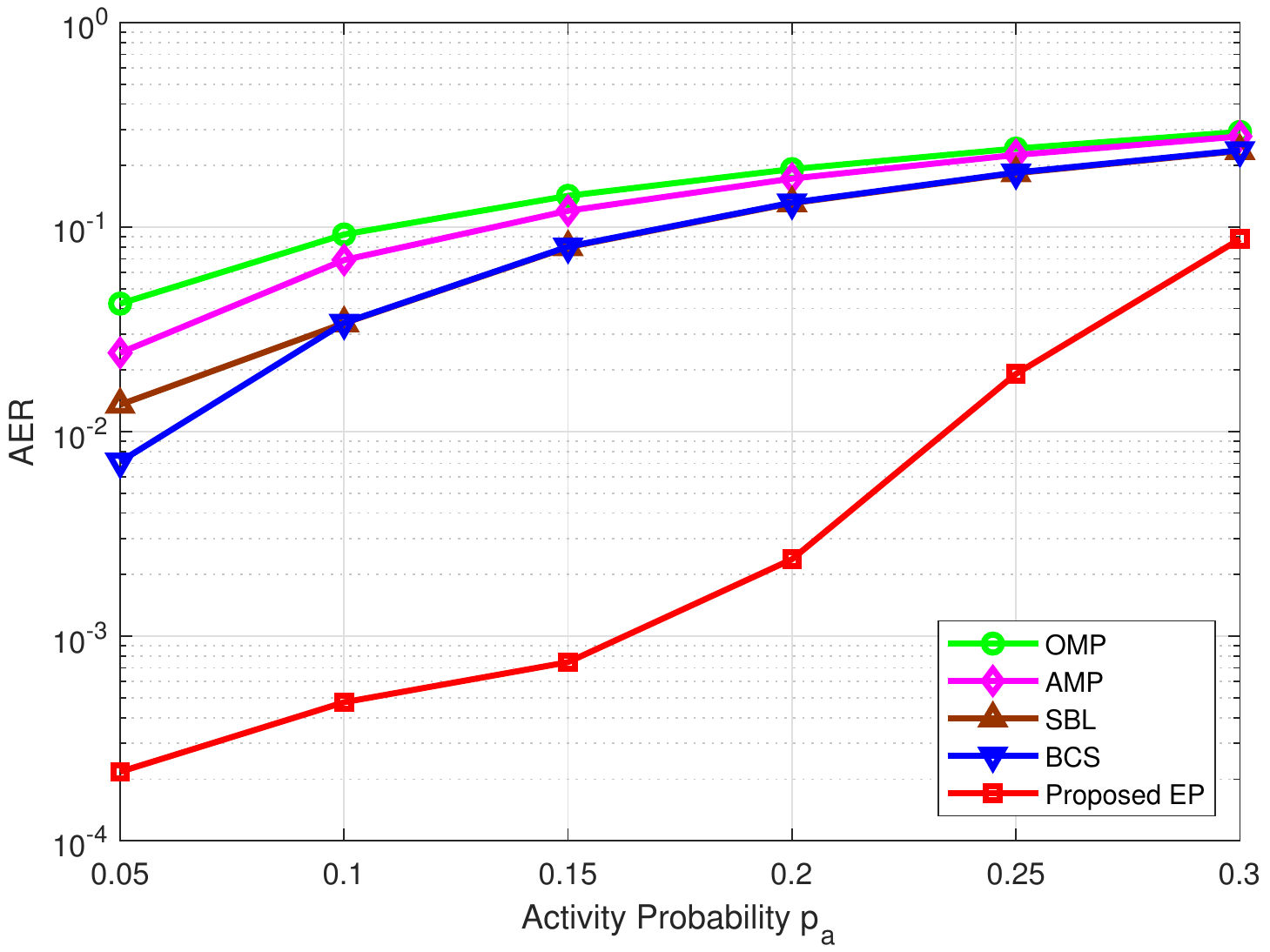}}\label{fig:sim_3_a}
		\vfill
		\subfloat[]{\includegraphics[width=0.8\linewidth]{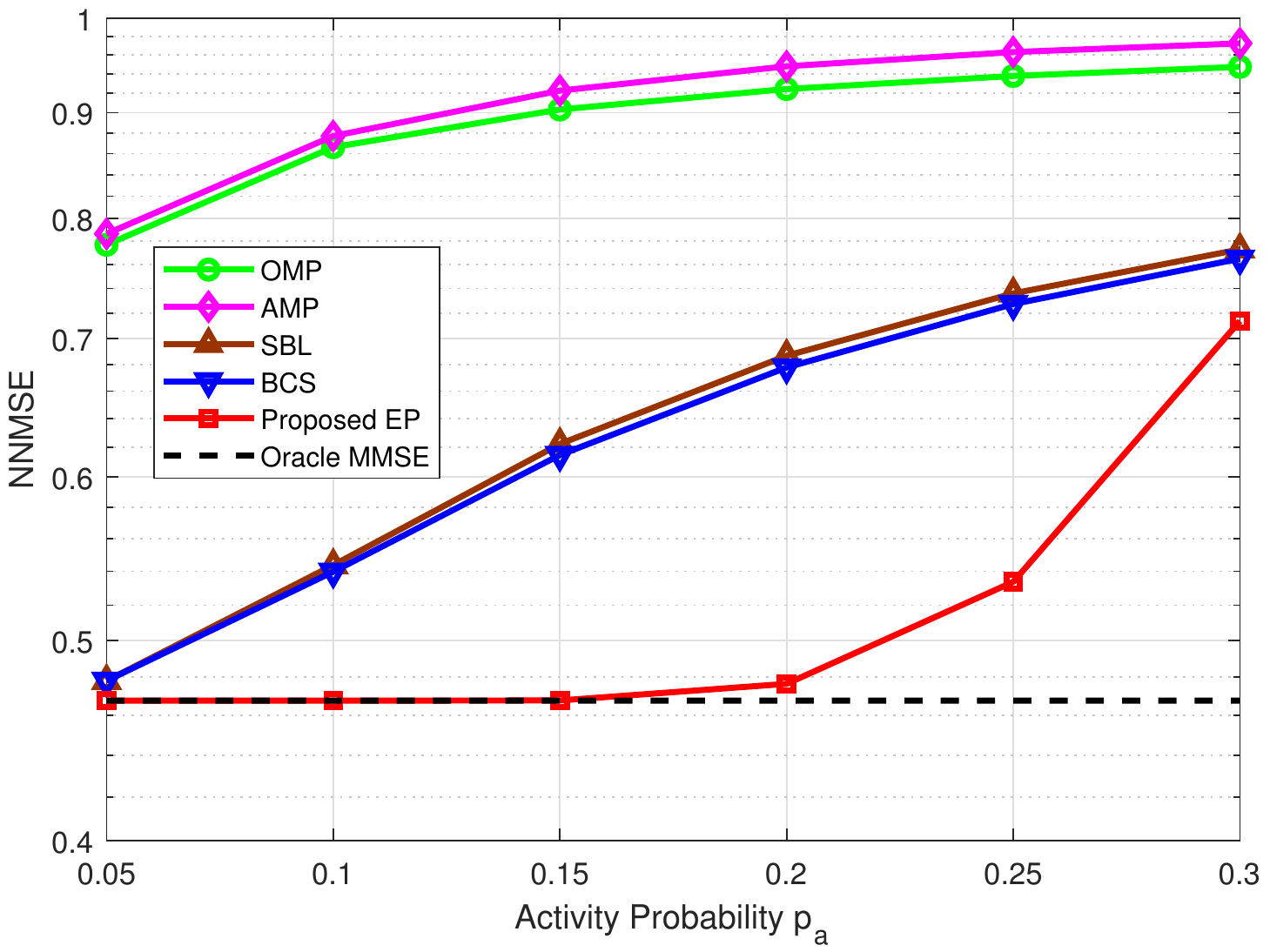}}\label{fig:sim_3_b}
	\end{figure}
	\begin{figure}[t]\ContinuedFloat
		\centering
		\subfloat[]{\includegraphics[width=0.8\linewidth]{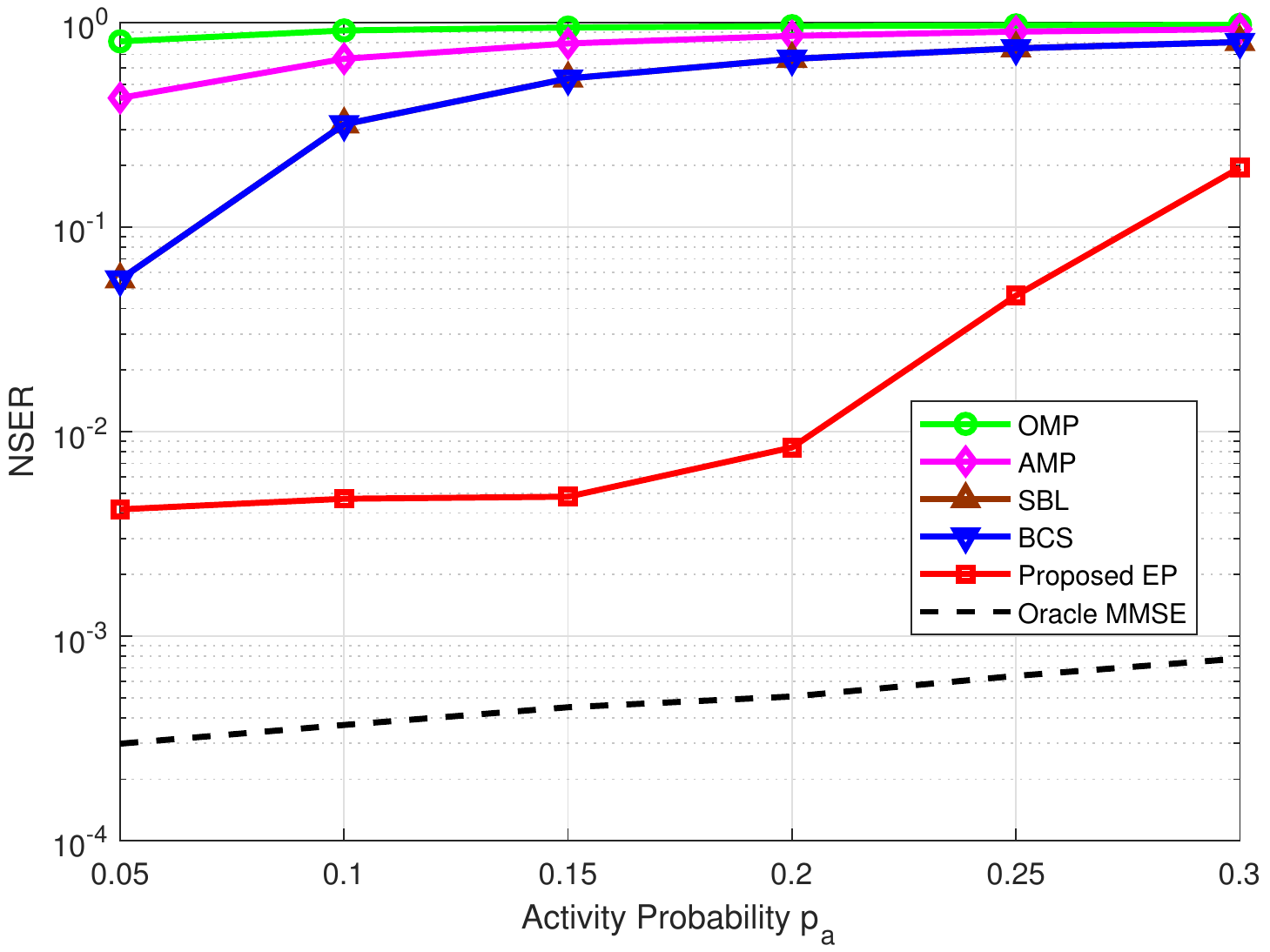}}
		\label{fig:sim_3_c}
		\addtocounter{figure}{+1}
		\caption{(a) AER, (b) NNMSE, and (c) NSER as a function of the activity 
			probability $p_a$.}
		\label{fig:sim_3}
	\end{figure}
	
	\clearpage
	\begin{figure}[t]
		\centering
		\includegraphics[width=0.8\linewidth]{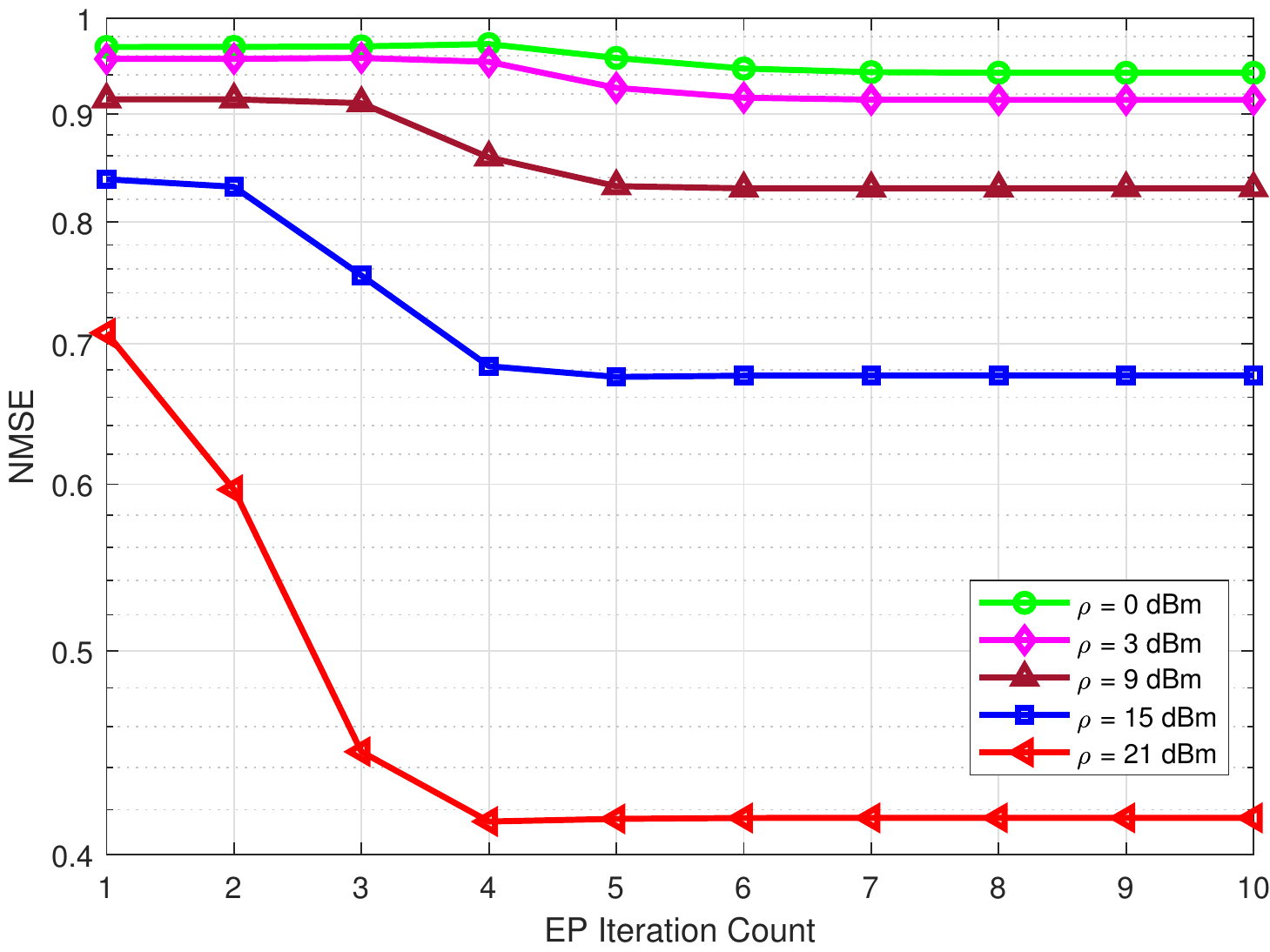}
		\caption{NMSE of the target vector $\mathbf{g}$ as a function of the EP iteration index $l$.}
		\label{fig:sim_5}
	\end{figure}

	\clearpage
	\subsection{Simulation Results}
	Fig.~\ref{fig:sim_1} shows the AER, NNMSE, and NSER performance when $p_a = 0.1$ and $M=64$. The non-Bayesian greedy algorithms such as OMP and AMP perform worse than the Bayesian algorithms because they do not rely on the statistical distributions of user activities and channels. SBL and BCS outperform the non-Bayesian greedy algorithms. However, since SBL and BCS exploit the distribution derived from the received signal (which is not necessarily correct), there is a substantial performance gap from the proposed technique. We observe that the proposed EP estimator outperforms the conventional algorithms. In particular, the proposed technique performs close to the Oracle MMSE estimator when the transmit power is larger than $12 \textrm{ dBm}$.
	
	Fig.~\ref{fig:sim_2} shows the performance when the length of spreading sequences $M$ varies from $32$ to $96$. In this simulation, we set the transmit power to $\rho = 20\textrm{ dBm}$. We can clearly observe that the performance degrades as $M$ decreases because the ratio $M/N$ of the system decreases (i.e., the system becomes more underdetermined). Note that the performance in the small $M/N$ regime is important for the massive connectivity scenario. Even when $M/N = 0.25$, the proposed technique achieves acceptable performance with the NSER of about $10^{-1}$.
	
	Fig.~\ref{fig:sim_3} shows the performance when the activity probability $p_a$ varies from $0.05$ to $0.3$. The transmit power and the length of spreading sequences are set to $\rho = 20\textrm{ dBm}$ and $M = 64$, respectively. We observe that all algorithms under test including the proposed algorithm are degraded when $p_a$ increases. This is because the interference among devices increases as more devices are active. The proposed algorithm outperforms OMP, AMP, SBL, and BCS by a large margin even if $p_a$ is higher than $0.2$. 
	
	In Fig.~\ref{fig:sim_5}, we investigate the convergence behavior of the proposed EP estimator. We plot the normalized mean squared error (NMSE) of the estimate of the target vector $\mathbf{g}$, i.e., $\frac{\|\hat{\mathbf{g}} - \mathbf{g}\|_2^2}{\|\mathbf{g}\|_2^2}$, as a function of the EP iteration index $l$. The smoothing parameter $\beta$ is set to $0.9$. As shown in Fig.~\ref{fig:sim_5}, we observe that the proposed EP estimator converges to the true solution within $4$ to $6$ iterations for all transmission power regimes. The iterative EP algorithm provides a reliable approximate distribution $q(\mathbf{g})$ whose mean is close to the mean of the target distribution $p(\mathbf{g})$ only with a few iterations.
	
	\clearpage
	\section{Concluding Remarks} 
	In this paper, we have proposed an EP-based Bayesian joint active user detection (AUD) and channel estimation (CE) technique for mMTC systems. Our work is motivated by the observation that most of conventional CS-MUD schemes are based on non-Bayesian approaches, and they cannot make the most of the statistical prior distribution of the user activities and channels. By exploiting the prior distribution, the proposed technique iteratively finds the best approximate Gaussian distribution that is close to the posterior distribution of the target composite vector of the user activities and channels. Then, by solving the MAP estimation problem with the approximate distribution, AUD and CE are performed jointly. The data detection (DD) of active devices are then performed based on the obtained knowledge of user activities and channels. From numerical simulations, we have demonstrated that the proposed technique achieves significant performance gains in terms of AUD, CE, and DD over the conventional sparse recovery algorithms.
	
	\clearpage
	\appendices
	\section{Derivations of The Moments Computations}
	The $m$-th ($m=0,1,2$) moments of the distribution $f_{2,n}q_{\backslash 2,n}^{(l)}$ with respect to $g_n$ are computed. First, the zeroth moment $G_{0,n}^{(l)}$ is computed as
	\begin{align}
		G_{0,n}^{(l)} &= \int_{-\infty}^{\infty} f_{2,n}(g_n)q_{\backslash 2,n}^{(l)}(g_n)\mathrm{d}g_n \nonumber\\
		&=\int_{-\infty}^{\infty} \left[(1-p_n)\delta(g_n)+p_n\mathcal{CN}(g_n|0,\alpha_n)\right]\cdot\mathcal{CN}(g_n|\tilde{m}_{\backslash 2,n}^{(l)},\tilde{v}_{\backslash 2,n}^{(l)})\mathrm{d}g_n \nonumber\\
		&=(1-p_{n})\int_{-\infty}^{\infty} \delta(g_n)\mathcal{CN}(g_n|\tilde{m}_{\backslash 2,n}^{(l)},\tilde{v}_{\backslash 2,n}^{(l)})\mathrm{d}g_n+p_{n}\int_{-\infty}^{\infty} \mathcal{CN}(g_n|0,\alpha_n)\mathcal{CN}(g_n|\tilde{m}_{\backslash 2,n}^{(l)},\tilde{v}_{\backslash 2,n}^{(l)})\mathrm{d}g_n \nonumber\\
		&=(1-p_{n})\mathcal{CN}(0|\tilde{m}_{\backslash 2,n}^{(l)},\tilde{v}_{\backslash 2,n}^{(l)})\mathrm{d}g_n+p_{n}\int_{-\infty}^{\infty} \mathcal{CN}(g_n|0,\alpha_n)\mathcal{CN}(g_n|\tilde{m}_{\backslash 2,n}^{(l)},\tilde{v}_{\backslash 2,n}^{(l)})\mathrm{d}g_n 
		\label{eqn:G_0}.
	\end{align}
	The product of two complex Gaussian PDFs, $p_1(g_n) = \mathcal{CN}(g_n|\mu_1,\sigma_1^2)$ and $p_2(g_n)=\mathcal{CN}(g_n|\mu_2,\sigma_2^2)$, is given by \cite{Bromiley03}
	\begin{align}
		p_1(g_n)p_2(g_n)&=\mathcal{CN}(\mu_2|\mu_1,\sigma_1^2+\sigma_2^2)\cdot\mathcal{CN}(g_n|\frac{\mu_1\sigma_2^2+\mu_2\sigma_1^2}{\sigma_1^2+\sigma_2^2},\frac{\sigma_1^2\sigma_2^2}{\sigma_1^2+\sigma_2^2}).
	\end{align}
	Using the above formula, \eqref{eqn:G_0} is rewritten as
	\begin{align}
		G_{0,n}^{(l)} &=(1-p_{n})\mathcal{CN}(0|\tilde{m}_{\backslash 2,n}^{(l)},\tilde{v}_{\backslash 2,n}^{(l)}) \nonumber\\
		&\quad+p_{n}\int_{-\infty}^{\infty}\mathcal{CN}(0|\tilde{m}_{\backslash 2,n}^{(l)},\alpha_n+\tilde{v}_{\backslash 2,n}^{(l)})\cdot\mathcal{CN}(g_n|\frac{\tilde{m}_{\backslash 2,n}^{(l)}\alpha_n}{\alpha_n+\tilde{v}_{\backslash 2,n}^{(l)}},\frac{\alpha_n\tilde{v}_{\backslash 2,n}^{(l)}}{\alpha_n+\tilde{v}_{\backslash 2,n}^{(l)}})\mathrm{d}g_n
	\end{align}
	
	We next compute the first moment $G_{1,n}^{(l)}$ as
	\begin{align}
	G_{1,n}^{(l)} &= \int_{-\infty}^{\infty} g_n f_{2,n}(g_n)q_{\backslash 2,n}^{(l)}(g_n)\mathrm{d}g_n \nonumber\\
	&=\int_{-\infty}^{\infty} g_n\left[(1-p_n)\delta(g_n)+p_n\mathcal{CN}(g_n|0,\alpha_n)\right]\cdot\mathcal{CN}(g_n|\tilde{m}_{\backslash 2,n}^{(l)},\tilde{v}_{\backslash 2,n}^{(l)})\mathrm{d}g_n \nonumber \\
	&=(1-p_{n})\int_{-\infty}^{\infty} g_n\delta(g_n)\mathcal{CN}(g_n|\tilde{m}_{\backslash 2,n}^{(l)},\tilde{v}_{\backslash 2,n}^{(l)})\mathrm{d}g_n \nonumber\\
	&\quad+p_{n}\int_{-\infty}^{\infty} g_n\mathcal{CN}(g_n|0,\alpha_n)\mathcal{CN}(g_n|\tilde{m}_{\backslash 2,n}^{(l)},\tilde{v}_{\backslash 2,n}^{(l)})\mathrm{d}g_n \nonumber\\
	&=p_{n}\int_{-\infty}^{\infty} g_n\mathcal{CN}(g_n|0,\alpha_n)\mathcal{CN}(g_n|\tilde{m}_{\backslash 2,n}^{(l)},\tilde{v}_{\backslash 2,n}^{(l)})\mathrm{d}g_n \nonumber\\
	&=p_{n}\mathcal{CN}(0|\tilde{m}_{\backslash 2,n}^{(l)},\alpha_n+\tilde{v}_{\backslash 2,n}^{(l)})\cdot\int_{-\infty}^{\infty}g_n\mathcal{CN}(g_n|\frac{\tilde{m}_{\backslash 2,n}^{(l)}\alpha_n}{\alpha_n+\tilde{v}_{\backslash 2,n}^{(l)}},\frac{\alpha_n\tilde{v}_{\backslash 2,n}^{(l)}}{\alpha_n+\tilde{v}_{\backslash 2,n}^{(l)}})\mathrm{d}g_n \nonumber\\
	&=p_{n}\mathcal{CN}(0|\tilde{m}_{\backslash 2,n}^{(l)},\alpha_n+\tilde{v}_{\backslash 2,n}^{(l)})\frac{\tilde{m}_{\backslash 2,n}^{(l)}\alpha_n}{\alpha_n+\tilde{v}_{\backslash 2,n}^{(l)}}.
	\end{align}
	
	Similarly, the second moment $G_{2,n}^{(l)}$ is computed as
	\begin{align}
	G_{2,n}^{(l)} &= \int_{-\infty}^{\infty} \left|g_n\right|^2 f_{2,n}(g_n)q_{\backslash 2,n}^{(l)}(g_n)\mathrm{d}g_n \nonumber\\
	&=\int_{-\infty}^{\infty} 
	\left|g_n\right|^2\left[(1-p_n)\delta(g_n)+p_n\mathcal{CN}(g_n|0,\alpha_n)\right]\cdot\mathcal{CN}(g_n|\tilde{m}_{\backslash 2,n}^{(l)},\tilde{v}_{\backslash 2,n}^{(l)})\mathrm{d}g_n \nonumber\\
	&=(1-p_{n})\int_{-\infty}^{\infty} \left|g_n\right|^2\delta(g_n) \mathcal{CN}(g_n|\tilde{m}_{\backslash 2,n}^{(l)},\tilde{v}_{\backslash 2,n}^{(l)})\mathrm{d}g_n \nonumber\\
	&\quad+p_{n}\int_{-\infty}^{\infty} \left|g_n\right|^2\mathcal{CN}(g_n|0,\alpha_n)\mathcal{CN}(g_n|\tilde{m}_{\backslash 2,n}^{(l)},\tilde{v}_{\backslash 2,n}^{(l)})\mathrm{d}g_n \nonumber\\
	&=p_{n}\int_{-\infty}^{\infty} \left|g_n\right|^2\mathcal{CN}(g_n|0,\alpha_n)\mathcal{CN}(g_n|\tilde{m}_{\backslash 2,n}^{(l)},\tilde{v}_{\backslash 2,n}^{(l)})\mathrm{d}g_n \nonumber\\
	&=p_{n}\mathcal{CN}(0|\tilde{m}_{\backslash 2,n}^{(l)},\alpha_n+\tilde{v}_{\backslash 2,n}^{(l)})\cdot\int_{-\infty}^{\infty}\left|g_n\right|^2\mathcal{CN}(g_n|\frac{\tilde{m}_{\backslash 	 2,n}^{(l)}\alpha_n}{\alpha_n+\tilde{v}_{\backslash 2,n}^{(l)}},\frac{\alpha_n\tilde{v}_{\backslash 2,n}^{(l)}}{\alpha_n+\tilde{v}_{\backslash 2,n}^{(l)}})\mathrm{d}g_n \nonumber\\
	&=p_{n}\mathcal{CN}(0|\tilde{m}_{\backslash 2,n}^{(l)},\alpha_n+\tilde{v}_{\backslash 2,n}^{(l)})\cdot\left(\left|\frac{\tilde{m}_{\backslash 2,n}^{(l)}\alpha_n}{\alpha_n+\tilde{v}_{\backslash 2,n}^{(l)}}\right|^2+\frac{\alpha_n\tilde{v}_{\backslash 2,n}^{(l)}}{\alpha_n+\tilde{v}_{\backslash 2,n}^{(l)}}\right).
	\end{align}
	
	
\end{document}